# Cosmological constraints from the Hubble diagram of quasars at high redshifts

G. Risaliti[1,2*], E. Lusso[3]

[1]Dipartimento di Fisica e Astronomia, Università degli Studi di Firenze, Via. G. Sansone 1, 50019 Sesto Fiorentino (FI), Italy.
[2]INAF- Osservatorio Astrofisico di Arcetri, Largo E. Fermi 5 50125 Firenze, Italy.
[3]Centre for Extragalactic Astronomy, Durham University, South Road, Durham, DH1 3LE, UK.

**The *concordance* (ΛCDM) model reproduces the main current cosmological observations[1,2,3,4] assuming the validity of general relativity at all scales and epochs, the presence of cold dark matter, and of a cosmological constant, Λ, equivalent to a *dark energy* with constant density in space and time. However, the ΛCDM model is poorly tested in the redshift interval between the farthest observed Type Ia supernovae[5] and that of the Cosmic Microwave background (CMB). We present new measurements of the expansion rate of the Universe in the range 0.5<$z$<5.5 based on a Hubble diagram of quasars. The quasar distances are estimated from their X-ray and ultraviolet emission, following a method developed by our group[6,7,8]. The distance modulus-redshift relation of quasars at $z$<1.4 is in agreement with that of supernovae and with the concordance model. Yet, a deviation from the ΛCDM model emerges at higher redshift, with a statistical significance of ~4σ. If an evolution of the dark energy equation of state is allowed, the data suggest a dark energy density increasing with time.**

Quasars are the most luminous persistent sources in the Universe, observed up to redshifts $z$=7.5[9,10]. According to a generally accepted scenario, the ultraviolet (UV) emission is produced by an accretion disk where the gravitational energy of the infalling material is partially transformed into radiation[11], while the X-rays are produced by a plasma of hot relativistic electrons (the so-called *corona*) around the accretion disk, through inverse-Compton scattering processes on the seed UV photons. A non-linear relation between these two components has been known for more than 30 years[12], and is usually parametrized as log($L_X$)=γlog($L_{UV}$)+β, where $L_X$ and $L_{UV}$ are the rest-frame monochromatic luminosities at 2 keV and at 2500 Å, respectively, and γ~0.6. The observed dispersion is of the order of δ~0.35-0.40 for several literature samples[13,14,15]. In the past few years we have discovered that most of the observed dispersion is not intrinsic, but it is rather due to observational effects[7]. We have gradually refined our selection technique and flux measurements, and we have estimated an intrinsic dispersion smaller than 0.15 dex. This discovery has two major implications, both of which may transform their respective fields. Firstly, in cosmology, our result provides a new standard candle, like Type Ia supernovae, but one that can be used to much higher redshifts and earlier times. Secondly, a strong regularity in the emission from powerful quasars, where previous ideas were of a highly stochastic nature, allow us to make progress in understanding the governing physics of gas accreting onto supermassive black holes. In this manuscript, we focus on two new major observational advances in this context. First, we produced a new sample of about 1,600 quasars with highly reliable X-ray and UV measurements from public archives, selected in order to minimize any known systematic effect on the X-ray to UV flux ratio. Second, we performed dedicated X-ray observations on a sample of 30 high-luminosity optically selected quasars in the redshift range $z$=3.0-3.3, within a Very Large Program with ESA's X-ray observatory *XMM-Newton*[16], significantly increasing the reliability of our cosmological analysis.

The bulk of the quasar parent sample has been obtained from the cross-correlation of the XMM-*Newton* Serendipitous Source Catalogue Data Release 7[17] with the *Sloan Digital Sky Survey* (SDSS) quasar catalogues from Data Releases 7[18] and 12[19]. The number of unique SDSS quasars with X-ray detection is 7,237 sources. Our main aim is to obtain a final quasar sample with reliable measurements of the *intrinsic* X-ray emission of the corona, and of the UV emission from the disc, avoiding any possible physical and/or observational contaminants that could bias these measurements. These contaminants include strong radio sources, intrinsic dust reddening in the UV, gas absorption in the X-rays, and selection effects due to the quasar X-ray and optical variability which would bias our data towards brighter states. The filtering process is summarized in the Methods and fully described in a forthcoming dedicated publication. By applying our filtering procedure, we are left with ~1,400 quasars with reliable measurements of both the X-ray and UV flux.

To better sample the Hubble Diagram at high redshifts, we included 38 quasars at $z$>4 where we reanalyzed the X-ray and UV spectroscopic data and 19 quasars from the *Champ* survey[20]. We also included a low redshift ($z$<0.2) sample of 18 active galactic nuclei with UV and X-ray data from *Swift*[21]. We further added 102 quasars from the *XMM-COSMOS*[14,22] survey not taken into account in the previous selection. Finally, we considered 18 quasars, out of a total sample of 30 sources, with pointed XMM-*Newton* observations from our Very Large Programme that fulfilled our filters.

The final quasar sample consists of ~1,600 quasars in the redshift range 0.04<$z$<5.1. This is a significant increase with respect to previous works, but only less than 20% of the total initial sample. This is a consequence of our conservative approach in the filtering procedure. Given that systematic effects in the flux measurements are potentially the main issue in the cosmological application, in this first stage of our work we choose the highest possible level of reliability over sample size and statistical significance (further details are given in the Methods and Supplementary Data).

The 30 new high-redshift quasars have been selected amongst the most luminous sources within the SDSS DR7[18] quasar catalogue in the redshift range $z$=3.0-3.3. The optical (rest-frame UV) spectra show very homogeneous properties, with no evidence of reddening/absorption and with a blue continuum emission, typical of radiatively efficient accretion disks[11]. The X-ray properties derived from the new observations reveal two clearly distinct groups of sources. About 70% of the sample show bright X-ray spectra, well described by a power law with photon index $\Gamma \sim 2$, whilst the remaining 30% is characterized by a much harder ($\Gamma < 1.7$) and fainter emission. In no cases we found evidence for significant gas absorption in excess of the Galactic value. The X-ray to UV ratio is close to the expected value from the X-ray to UV relation calibrated in our previous works[7,8] for the X-ray bright group, while it is lower by up to a factor of 20 for the faint group (Supplementary Figure 6). These results are relevant in several respects. The two groups may point to two different physical states of the X-ray emitting region, without any significant difference in the physical properties of the disk (i.e. similar ionizing spectral slopes). The bright sample show a dispersion in the UV to X-ray luminosity relation of only 0.12 dex, which is significantly lower than the dispersion obtained in all our previous studies, demonstrating that most of the observed dispersion is due to observational issues rather than to intrinsic effects. Furthermore, considering that the intrinsic X-ray and UV variability of highly luminous quasars is of the order of 0.05-0.08 dex in log units (~10-20%) and 0.04 dex (~10%), respectively[23,24], and that the disk orientation with respect to the line of sight must add some dispersion to the observed relation, we conclude that the *intrinsic* dispersion of the X-ray to UV relation may not higher than 0.08 dex. On the other hand, the X-ray faint group is strongly offset from the relation. Our analysis demonstrates that the X-ray to UV relation holds only for UV blues, X-ray soft quasars, i.e. objects with both the accretion disk and the X-ray corona in a radiatively efficient state. Finally, the difference in X-ray spectral slope values between the two groups of quasars is a viable method to select the sources with the "right" X-ray properties for the analysis of the relation as well as for its cosmological application. As a result of that, only sources with $\Gamma > 1.7$ have been included in the final SDSS quasar sample considered in our analysis.

The homogenous quasar sample described above has been used to build the distance-luminosity relation (i.e. the *Hubble diagram*, see Figure 2), following the prescription described in our previous works[6,7]. Briefly, the log-linear relation, $\log(L_X) = \gamma \log(L_{UV}) + \beta$, shown in Figure 1 for the parent and the clean sample of quasars, can be expressed in terms of the rest-frame X-ray and UV fluxes, $F_X$ and $F_{UV}$, and the luminosity distance ($D_L$) of each source. The latter parameter can be then derived from the fluxes, as a function of the slope $\gamma$ and the normalization $\beta$ of the relation. The parameter $\gamma$ can be obtained in a cosmology-independent way following the procedure outlined in the Methods, based on fitting the $F_X$ - $F_{UV}$ relation in narrow redshift intervals. In this way, it is possible to test the non-evolution of the relation with the redshift, which is a necessary condition for its implementation to derive quasar distances. The parameter $\beta$ is an absolute scaling and cannot be determined independently from external calibrators. We estimated such a scaling by cross-matching the Hubble diagram of quasars with that of supernovae[2] in the common redshift range $z$=0.1-1.4, as shown in the zoom-in of Figure 2.

The results, shown in Figures 2 and 3, confirm the distance-redshift relation obtained with supernovae at redshift $z$<1.4. A fit with the concordance flat ΛCDM model provides a best fit parameter $\Omega_M$=0.31±0.05 (while $\Omega_\Lambda$ is fixed to 1-$\Omega_M$), in agreement with all the main current cosmological probes. Yet, the Hubble diagram of quasars show a significant deviation from the extrapolation of the ΛCDM model in the poorly explored redshift range $z$>1.4. To quantify this discrepancy, we adopted a *cosmographic* approach by fitting the Hubble diagram with a polynomial function of the quantity log(1+$z$): $D_L = k \sum_i a_i [\log(1+z)]^i$. The coefficient, $k$, and the first term of this polynomial expansion, $a_1$, are set to be equal to ln(10)c/$H_0$ and 1, respectively, to reproduce the Hubble law at low redshifts. This is a variation of the standard cosmographic approach based on a polynomial expansion in terms of the redshift $z$, with the advantage of a faster convergence at $z$>1 (Supplementary Figure 2). We obtained an excellent fit of the combined supernovae and quasars Hubble diagram with the first three terms of the polynomial function, i.e. with two free parameters, $a_2$ and $a_3$. The dispersion around the best fit relation is smaller than for all the cosmological models we investigated, and we found no structure in the residuals (Supplementary Figure 5). By assuming a flat ΛCDM model, with simple algebra, it is possible to derive the parameters $a_2$ and $a_3$ as a function of the cosmological parameter $\Omega_M$ (see Methods for details). From $a_2(\Omega_M)$ and $a_3(\Omega_M)$, we can then derive a relation between $a_2$ and $a_3$, which must hold within a flat ΛCDM model. The comparison between the values of $a_2$ and $a_3$ obtained from the fit and this relation represent a model-independent, *cosmographic* test of the flat ΛCDM model. The results plotted in Figure 3 show a discrepancy between the ΛCDM model and the data-derived coefficients which is statistically significant at the 4σ level. This comparison method can be in principle extended to any cosmological model. It is beyond the purposes of this work to analyze the possible extensions of the standard cosmological model that best fit the data. Here we only show two tests of the most common extensions of the ΛCDM model. When the data are fitted with a flat cosmological model with a free value of the equation of state parameter $w$ ($w$=-1 corresponds to a cosmological constant), the best fit is obtained with a high value of the matter density ($\Omega_M$>0.4) and a value of the dark energy parameter $w$<-1, i.e. a dark energy density increasing with time (Figure 3). If the dark energy equation of state is allowed to vary with redshift, adopting the usual parametrization $w=w_0+w_a(1-a)$, where $a=1/(1+z)$ is the scale factor of the Universe, we obtain loose constraints on the parameters $w_0$ and $w_a$ from the Hubble diagram alone. Figure 4 shows the constraints on $w_0$-$w_a$ when we combine our results with those from the Cosmic Microwave Background and weak lensing, finding values consistent with $w$<-1.3. This kind of solution is also one of the possible ways to solve the current 3.7σ tension between the values of the Hubble constant derived from local indicators[25] and from the CMB[1,26]. We note that the overlap between CMB + weak lensing (green contours) and quasars + SNe (orange/red contours) occurs in a relatively narrow region of the parameter space. As already discussed by the Planck collaboration[1,26], contours' overlap in the marginalized probability distributions does not guarantee consistency in the full parameter space. More detailed studies of the cosmological implications of the Hubble diagram, and on the comparison with other recent relevant measurements, such as the new z>1.5 supernovae[27] and the Baryonic Acoustic Oscillation at z~2.3, will follow in forthcoming publications.

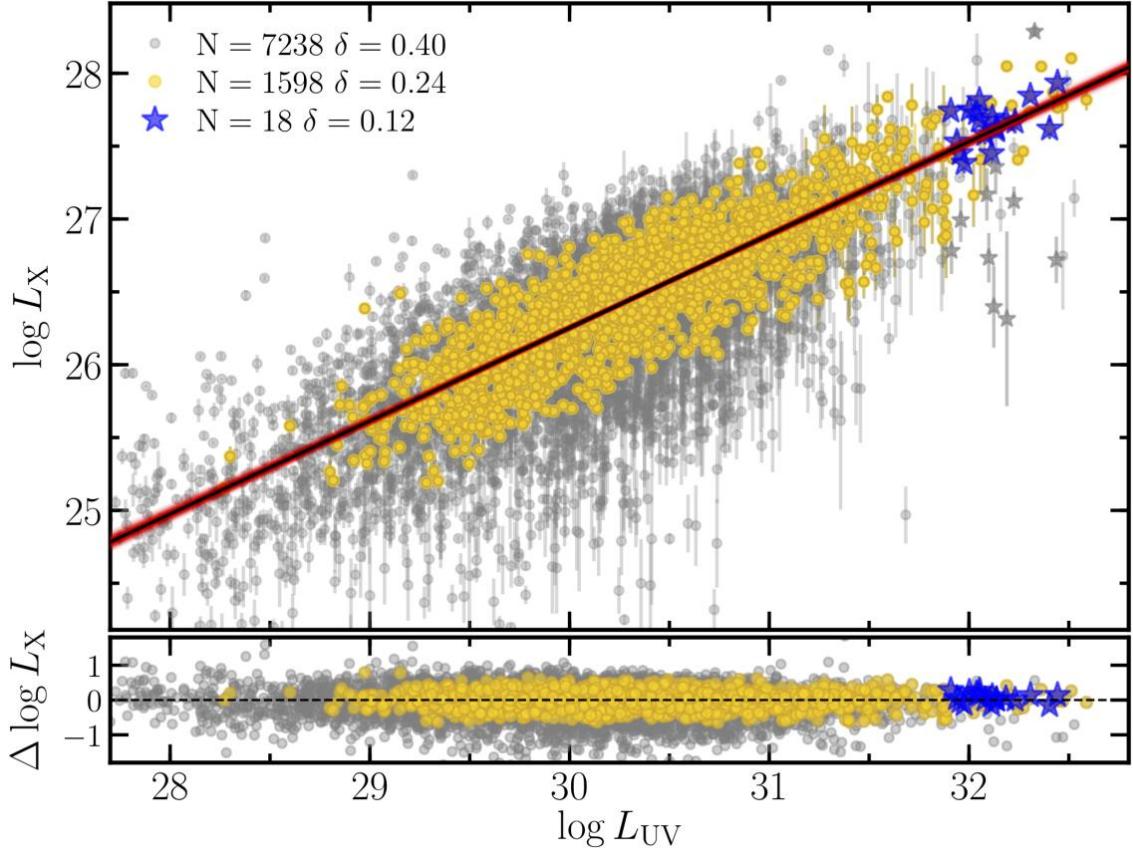

**Figure 1:** The rest-frame monochromatic UV to X-ray luminosity (in erg/s/cm$^2$/Hz) relation for the whole sample of ~7,300 quasars with available X-ray and UV measurements (grey points), and for the ~1,600 quasars that fulfill our selection criteria for the Hubble Diagram analysis (yellow points). The new sample of *z*>3 quasars with dedicated XMM-*Newton* observation is also shown with blue stars. The luminosities for this plot have been obtained from the measured fluxes assuming a standard ΛCDM cosmological model with $\Omega_M$=0.3, $\Omega_\Lambda$=0.7, and $H_0$=70 km/s/Mpc. The scatter, δ, on the UV to X-ray luminosity relation for the different quasar samples is also reported. The red solid line presents the best-fit regression with slope γ=0.633±0.002. The lower panel shows the residuals, Δ(log $L_X$), with respect to the best-fit regression line.

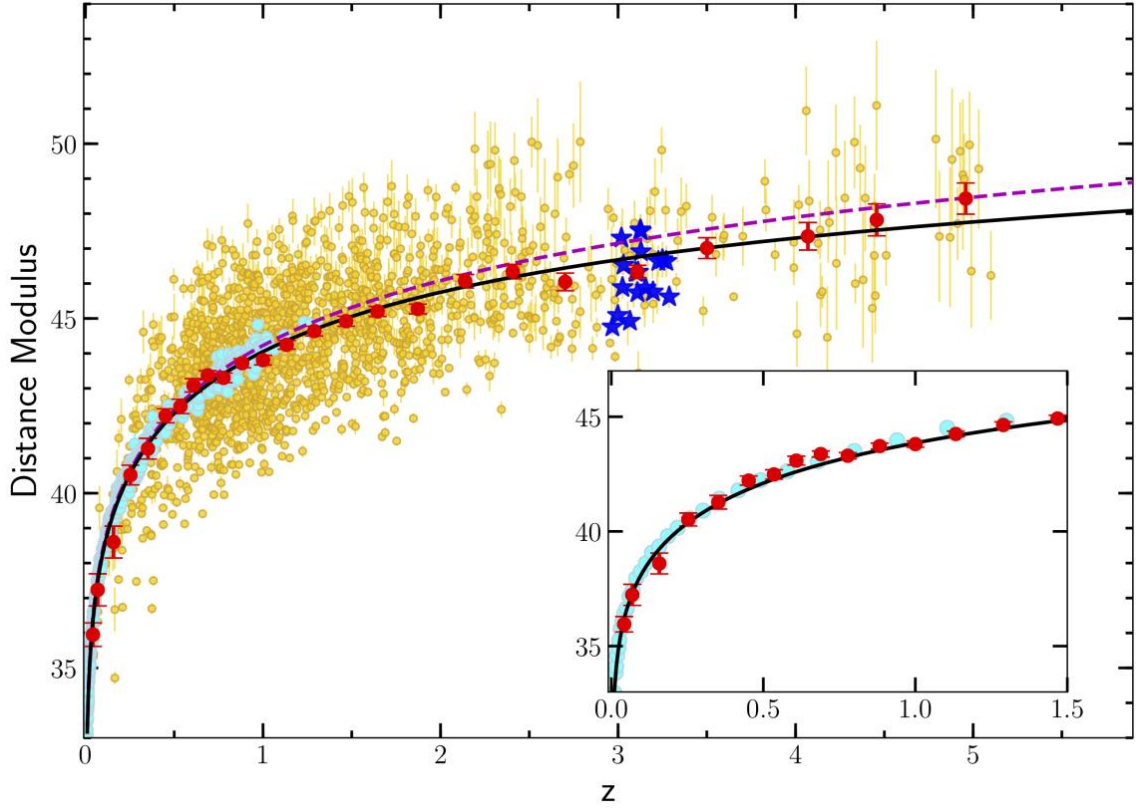

**Figure 2:** Hubble diagram of supernovae from the JLA survey[2] (cyan points) and quasars (yellow points). Red points represent the mean (and uncertainties on the mean) of the distance modulus in narrow redshift bins for quasars only. These averages are shown just for visualization and, as such, are not considered in the statistical analysis. The new sample of $z>3$ quasars with dedicated XMM-*Newton* observation is shown with blue stars. The inset is a zoom of quasar and supernovae averages in the common redshift range. The dashed magenta line shows a flat $\Lambda$CDM model with $\Omega_M=0.31\pm0.05$ fitting the $z<1.4$ data and extrapolated to higher redshifts. The black solid line is the best MCMC fit of the third order expansion of $\log(1+z)$.

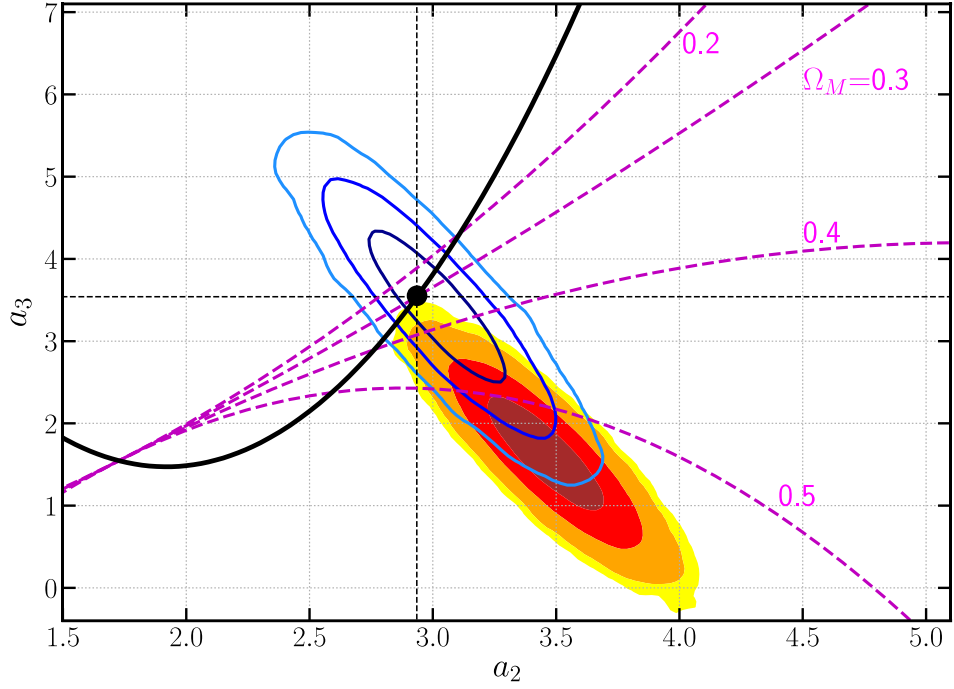

**Figure 3:** Comparison between the polynomial expansion P[log(1+z)] of the luminosity distance-redshift relation and the flat ΛCDM model, where $a_2$ and $a_3$ are the coefficients of the second and third term of P[log(1+z)], respectively. The black curve shows the relation between the two parameters in the flat ΛCDM model, described in the Methods. The black dot marks the value for $\Omega_M=0.3$. The filled contours are obtained from the fit of the third order polynomial P[log(1+z)] of the combined quasar and supernove Hubble diagram. The empty contours are obtained from the same fit to the data at $z<1.4$. Contour levels are at 1, 2, and 3σ. The dashed magenta lines are an example of the application of this cosmographic comparison to other cosmological models: we assumed a flat wCDM model, i.e. with a free dark energy density parameter $w$ ($w$=-1 corresponds to a cosmological constant), and we computed the expected values of $a_2$ and $a_3$ as a function of $w$ for different values of $\Omega_M$, as indicated in the labels. The intersections between the magenta lines and the black curve are the points with w=1 (left) and w=-1 (right), and the values of w decrease from left to right. While the concordance ΛCDM model is in agreement with the observational results for $z<1.4$, it is in tension with the whole data set at a statistical significance of 99.98%, corresponding to ~4σ. Within the wCDM model, the preferred values for the parameters are $\Omega_M>0.3$ and $w<-1.3$.

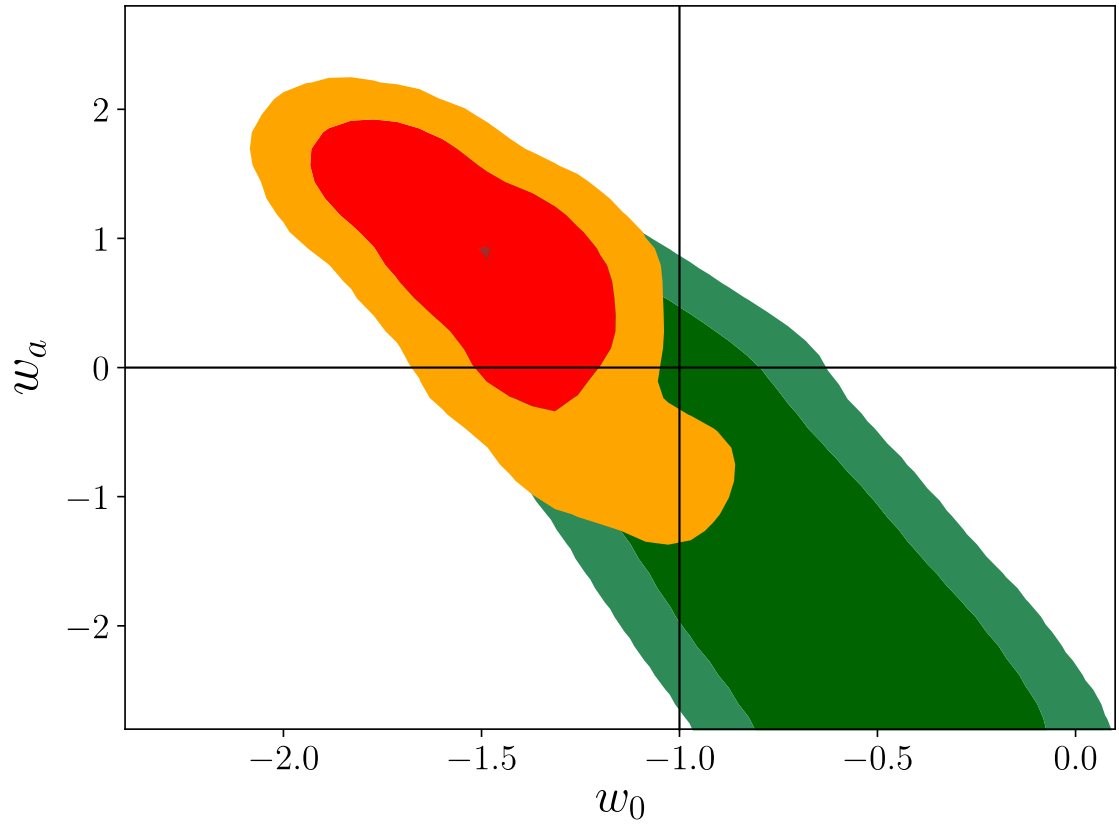

**Figure 4:** The *w₀-wₐ* error contours (2σ and 3σ) in a *w₀wₐ*CDM model from *Planck*+Weak Lensing data[1] (green) and from the combination of the same data with the Hubble diagram of supernovae and quasars (red-orange).

# Methods

In the main paper we presented a Hubble Diagram of 1,598 quasars, covering the redshift range z=0.036-5.100, and showing a 4-$\sigma$ deviation from the standard $\Lambda$CDM model at z>1.5. Here we give a technical description of our work, focusing on (1) how we built the parent quasar sample; (2) the selection of the final "clean" sample; (3) the creation and analysis of the Hubble diagram of quasars; and (4) the main checks on possible physical and observational biases. A fully detailed description of the latter point is presented in the Supplementary Data.

**1. Data set.** The sample used to build the Hubble diagram has been obtained from a parent sample of 7,237 sources. Here we describe its composition and the cleaning process to select the final 1,598 sources. The main sample is built merging five groups of quasars:

- <u>SDSS/XMM sample</u>. The bulk of the parent quasar sample has been obtained by cross-correlating the SDSS DR7[18] and DR12[19] quasar samples with the XMM-*Newton* Serendipitous Source Catalogue Seventh Data Release (3XMM-DR7)[17] of X-ray detections. We used a 3'' matching radius where broad absorption line quasars[27] and powerful radio sources[28-37] are excluded from the optical catalogues, as well as spurious X-ray detections with high level of background.

The rest-frame monochromatic fluxes at 2 keV, $F_X$, along with their uncertainties and photon index measurements are obtained from the observed soft (0.5-2 keV) and hard (2-12 keV) fluxes listed in the 3XMM-DR7 catalogue. We assumed a power law spectrum and adopted a custom procedure, described in the Supplementary Data, in order to improve the precision of the flux estimates.

The rest-frame monochromatic fluxes at 2500 Å, $F_{UV}$, for the SDSS-DR7 quasars have been obtained from the Shen et al. (2011) quasar catalogue[18]. For the SDSS-DR12 quasars, the $F_{UV}$ values have been estimated by fitting the multi-wavelength broad-band photometry provided in the catalogue. We applied a redshift-dependent correction obtained from a comparison between the spectroscopic estimates of Shen et al. with our photometric fits applied to the DR7 sample. The details of the procedure are described in the Supplementary Data. The final SDSS DR7+DR12 quasar sample is composed by 6,647 quasars, i.e. 2,226 and 4,421 quasars from SDSS DR7 and DR12, respectively. We prioritized SDSS-DR7 quasars for the overlapping objects, as we considered the rest-frame 2500Å flux measurements in the Shen et al quasar catalogue as the best measurements of the quasar intrinsic continuum.

- <u>XMM-COSMOS sample.</u> We included 466 quasars from the XMM-COSMOS survey[14,22,29,30]. Both UV and X-ray flux values are published by Lusso et al. (2010)[14].

- <u>High-redshift sample.</u> In order to increase the statistics at high redshift we included a sample of 24 quasars at *z*>3, unobscured in both optical and X-rays and with spectroscopic redshift, observed within the Chandra *Champ* survey[20]. We also included a sample of 54 quasars in the redshift range 3.99<*z*<7.08, with optical information available from the literature and publicly available X-ray observations with either XMM-*Newton* (whose archival X-ray data were not included in the 3XMM-DR7 catalogue) or *Chandra*[31]. The X-ray observations for the latter quasar sample have been fully reduced and analyzed by our group. The results will be presented in a forthcoming, dedicated paper.

- <u>Low-redshift sample.</u> We included a sample of low-redshift ($0.036<z<0.202$) radio quiet quasars (18 objects) with simultaneous UV and X-ray observations from *Swift*, which have been thoroughly investigated within the study of the UV to X-ray luminosity relation[21].

- <u>New z~3 sample.</u> We added a sample of 28 highly luminous ($L_{bol}$~$10^{47}$ erg/s) quasars in the redshift range $3.00<z<3.296$ with new X-ray observations obtained within our XMM-*Newton* Very Large Program (PI. Risaliti, Proposal ID: 080395). The full proposed sample is composed by 30 very bright quasars in a narrow redshift range centred at z~3.1 selected from SDSS-DR7. The XMM-*Newton* observations were carried out during the year 2017. One quasar has not been observed yet, and we further excluded another target which was not marked as radio loud in the SDSS DR7 catalogue but turned out to be radio loud. The exposure times vary from 28 ks to 46 ks, and the data for the 28 remaining targets have been fully reduced and analyzed following the standard procedure outlined in the XMM-*Newton* website. The X-ray spectra have been fitted with a model including a power law and the Galactic absorption. We found no cases where extra components, such as further absorption, emission lines, reflection components, were needed. The full analysis of the UV and X-ray spectral properties derived from the SDSS and XMM-*Newton* data, respectively, will be presented in a series of dedicated papers. Supplementary Figure 1 shows the distribution of the X-ray photon indexes as a function of the X-ray flux obtained from spectral fitting of the XMM-*Newton* data from our programme. The dashed line represents the selection adopted to define the clean quasar sample (see Section 2 of the Methods and Section 6 in the Supplementary Data for a more in depth discussion).

**2. "Cleaning" criteria.** In order to remove possible systematics, we applied several filters, which reduced the sample to less than 20% of its initial size.

- <u>*X-ray absorption.*</u> Photoelectric absorption can be present in a significant fraction of quasars optically classified as unobscured[38]. The distribution of the photon index in bright, unobscured quasars is peaked at $\Gamma=1.9$-$2.0$[39]. If gas absorption is present, and the X-ray spectrum is reproduced by a simple power law, the *observed* photon index $\Gamma$ becomes smaller. An *intrinsic* photon index $\Gamma<1.5$ is rare, and in most cases is due to gas absorption. Intrinsic values $1.5<\Gamma<1.7$ are observed in a minority of quasars, but the chance of the presence of some low-energy absorption is not negligible. Therefore, we conservatively opted to filter out from the final sample all the sources with $\Gamma>1.7$.

- <u>*Observational contaminants in the UV*.</u> Intrinsic UV extinction and host-galaxy contamination are two serious issues for our analysis, because they may introduce strong systematics, as well as increasing the average dispersion in the X-ray/UV relation. To minimize these contaminants, we defined three indicators of possible extinction: the slope $\Gamma_1$ of a log($\nu$)-log($F_\nu$) power law in the 0.3-1 μm (rest frame) range, the analogous slope $\Gamma_2$ in the 0.145-0.3 µm range (rest frame), and the Δ(g-i) color, defined as the difference between the observed (g-i) color and its median value in small redshift bins. We then required the data to fulfil the following criteria: $\Gamma_1$ >0.8, $\Gamma_2$ >0.4, and Δ(g-i)<0.1, which correspond to limiting the intrinsic extinction to E(B-V)<0.1 and the host galaxy contamination to be less than 10% at the rest frame 2500 Å. For the few objects in the low- and high-redshift subsamples (<10 % of the main sample) without spectra, we constructed multi-wavelength broad-band SEDs based on all the available photometry (e.g. 2MASS, WISE, GALEX), and we then applied equivalent filters based on the $\Gamma_1$ and $\Gamma_2$ obtained from the interpolated SED.

- *Eddington bias*. Quasars with an average X-ray flux close to the flux limit of a given observation will be detected only in the case of a positive flux fluctuation with respect to that limit. To avoid such a bias, we adopted a procedure which excludes from the sample all sources with an *expected* X-ray flux lower than a given threshold depending on their flux limit[6,7].

For our technique to be robust, we need to estimate the X-ray flux limit for each XMM observation. For the SDSS-3XMM quasar samples we estimated the X-ray flux limit for each PN observation in the 3XMM catalogue given the on-time exposure and off axis angle as described by Lusso & Risaliti (2016)[7]. We estimated the minimum detectable flux in the soft band as a function of the exposure time following the relations plotted in Figure 3 of Watson et al. (2001)[40] for the PN, taking into account the vignetting factor as a function of their respective off-axis values. For the XMM-COSMOS sample, we considered the survey flux limit in the soft band (i.e. $5\times10^{-16}$ erg/s/cm$^2$). The other subsamples are all pointed observations.

Our procedure is independent from the *actually observed* flux. However, in order to predict the *expected* flux value we need to assume the X-ray to UV luminosity relation. In principle, this may introduce a bias in the subsequent computation of the Hubble diagram of quasars, as we are using luminosities derived from the observed fluxes assuming a specific cosmological model. Nonetheless, if we use a conservative threshold, we expect that the fits to the Hubble Diagram will be independent from the cosmological model chosen in the filtering procedure. To demonstrate this point, and to estimate the optimal value of the flux threshold, we simulated several sets of mock samples of 100,000 quasars each, having the same UV and X-ray flux distributions, as well as the X-ray flux limit and redshift distributions, similar to our real sample. The X-ray fluxes have been derived from the UV to X-ray luminosity relation, adding a Gaussian intrinsic dispersion $\delta$=0.15 dex. The UV and X-ray luminosities in each mock sample have been then computed by assuming a different cosmological model. We generated a different quasar mock sample for all possible combinations of the cosmological parameters with both $\Omega_M$ and $\Omega_\Lambda$ ranging from 0.1 to 0.9 with a step of 0.2. We also created mock quasars samples with $\Omega_M$=0.3 and $\Omega_\Lambda$=0.7 and the *w* parameter fixed to 0.5 and -1.5. Finally, we investigated the case of $\Omega_M$=0.3 and $\Omega_\Lambda$=0.7 and *$w_0$*, *$w_a$* equal to -1.5 and 1, respectively, within the *$w_z$*CDM framework. This leads to more than 30 different mock quasar samples for which we studied the effect of the Malmquist bias given our approach detailed below.

For each mock quasar sample we assumed an X-ray flux limit (in monochromatic units), $F_{MIN}$, and we removed all sources with a simulated flux $\log F_{SIM} < \log F_{MIN} + k\delta$, where $k$ is a free coefficient, and $\delta$ is the dispersion added to each mock sample. The $F_{MIN}$ parameter is chosen to exclude about 20-25% of the sources (for $k=0$), which is close to the expected fraction of undetected objects, as discussed in Lusso & Risaliti 2016[7]. We finally fit the $L_X$-$L_{UV}$ relation (where the luminosities are obtained from the fluxes and the assumed cosmology) for the remaining objects. We repeated the procedure for different values of the parameter $k$, and we found that, for $k=2$ (i.e. choosing observations with a flux limit higher than the expected X-ray flux by at least twice the dispersion), we always recover the assumed $\Omega_M$ and $\Omega_\Lambda$. In the case of $k>2$, our results do not change, but the number of quasars in the final sample decreases, while for $k<2$ we start to see some small (on the order of a few percent) deviations from the assumed cosmological parameters. In particular, the value obtained for the slope of the $L_X$-$L_{UV}$ relation becomes flatter than the simulated one for $k<2$. When this procedure is applied to the real data, we recover the assumed slope ($\gamma=0.6$) for every $k>1$. This is probably due to the smaller size of the real sample, which make it insensitive to the small systematic effects present when $1<k<2$. In order to be conservative, we always adopt $k=2$ in the analysis of the real data and a representative dispersion of 0.24 dex.

The final result of the selection consists of 1,598 quasars: 791 sources from the SDSS-DR7 sample, 612 from the SDSS-DR12, 102 from XMM-COSMOS, 18 from the low-redshift sample, 19 from Chandra-*Champ*, 38 from the high-$z$ ($z>4$) sample, and 18 quasars from the new $z\sim3$ sample. The distribution of rest frame monochromatic luminosities at 2 keV as a function of redshift, for both the parent quasar sample and the clean one, is presented in the Extended Figure 2. A summary of the various AGN sub-samples for a given selection criteria is provided in the Extended Figure 3.

*Fitting algorithm.* The fit presented in the manuscript are all obtained, unless stated otherwise, by employing a fully Bayesian Monte Carlo Markov Chain (MCMC) based approach. Specifically, we employed the affine-invariance MCMC Python package *emcee*[41], where the estimated uncertainties on the parameters are taken into account. For our tests on censored data, and to investigate whether the fitting results depend on the adopted fitting algorithm, we also considered the LINMIX_ERR method[42], which is argued to be among the most robust regression algorithms with the possibility of reliable estimation of intrinsic random scatter on the regression. LINMIX_ERR accounts for measurement uncertainties on both independent and dependent variable, non-detections, and intrinsic scatter by adopting a Bayesian approach to compute the posterior probability distribution of parameters, given observed data.

**3. Analysis of the Hubble diagram of quasars.** The distance modulus-redshift relation for the quasar sample has been constructed with a two-steps procedure. First, we split the quasar sample in subsamples within narrow redshift bins ($\Delta \log z=0.05$), in order to have a spread in distance within each bin much smaller than the measured dispersion of the $L_{UV}$-$L_X$ relation. In these conditions, we can use fluxes as proxies of luminosities, and fit a log-linear $F_{UV}$-$F_X$ relation in each redshift bin in the form $\log(F_X)=\gamma\log(F_{UV})+\beta$ (see Supplementary Figure 4 for an example at $z\sim3$). In this way we achieve two goals: we test the fundamental physical assumption in our work of a non-evolution of the $L_{UV}$-$L_X$ relation with redshift, and we derive an average value for the slope $\gamma$. We repeated the analysis by varying the number of bins and their width in redshift, finding that the final value of $\gamma$ is insensitive to the specific choice of the redshift bins, provided that the redshift interval is $\Delta\log z<0.1$. Second, we derived a luminosity distance for each quasar, as a function of $F_X$, $F_{UV}$, the average slope $\gamma$ obtained by fitting the $F_{UV}$-$F_X$ relation in narrow redshift intervals, and the free parameter $\beta$:

$$\log D_L = \frac{1}{2-2\gamma}(\gamma \log F_{UV} - \log F_X) + \beta. \qquad (1)$$

To estimate the normalization of $\beta$, an external calibrator is needed. This has been done by cross-matching the JLA supernovae Ia sample with the quasar sample in the overlapping redshift range $z=0.1$-$1.4$.

Cosmological fits. The quasars and JLA samples have been fitted simultaneously with a *cosmographic* model consisting of a polynomial expansion of $\log(1+z)$: $D_L=k \Sigma_i a_i[\log(1+z)]^i$. The coefficient $k$ is set to be equal to $\ln(10)c/H_0$ whilst $a_1$ is fixed to 1 in order to reproduce the Hubble law locally, i.e. $D_L=c/H_0 \ln(1+z)$, where $H_0$ is fixed to 70 km/s/Mpc in our cosmographic fit of the data. The function we considered for our fit is the third order polynomial expansion, i.e. $D_L=\ln(10)c/H_0 (x + a_2 x^2 + a_3 x^3)$, where $x=\log(1+z)$. The distance modulus, DM, is thus defined as $DM=m+5[\log(D_L/Mpc)+5]$, where $m$, $a_2$, and $a_3$ are the free parameters. The parameter $m$ depends on the deviation of $H_0$ from the chosen reference value of 70 km/s/Mpc, and is very close to zero. A logarithmic expansion has the advantage of a more rapid convergence at high redshifts with respect to the standard linear expansion. This is illustrated in Extended Figure 5, where we show, as an example, the ratio between the distance-redshift relation obtained from a $\Lambda$CDM model with $\Omega_\Lambda=0.7$, $\Omega_M=0.3$, and the polynomial expansions in $z$ and $\log(1+z)$. We obtained a 1, 2 and $3\sigma$ constrains of the $a_2$ and $a_3$ parameters with an expansion to the third order in $\log(1+z)$ as shown in Extended Figure 5. Within the flat $\Lambda$CDM model, we can derive the relations between the coefficients of the polynomial expansion and $\Omega_M$ as: $a_2 = 3/2 - 3/4 \, \Omega_M$ and $a_3 = 7/6 - 2\Omega_M + 9/8 \, \Omega_M$. The blue curve in the Extended Figure 5 is then given by: $a_3 = 5/3 - 10/3 \, a_2 + 2a_2^2$. Adding a fourth order does not improve the fit, and the coefficient $a_4$ is poorly constrained. The black line shown in Figure 2 of the main paper is obtained by fitting the data with the third order polynomial expansion. The dispersion in the distance modulus-redshift relation is 1.4 dex, which corresponds to a dispersion in the $L_{UV}$-$L_X$ relation of 0.24 dex (see Supplementary Figure 6), based on Eq. 1.

**4. Tests of the results.** The new data presented in this work provide a precise estimate of the expansion of the Universe at z>2 and show a ~4σ tension with the predictions of the concordance model. Unexpected and/or radically new results require particularly accurate tests in order to rule out any observational bias. For this reason, we analysed the dependence of our results on all the main selection parameters (e.g. dust reddening, gas absorption, Eddington bias). We notice that the effects we need to investigate are primarily those with a possible dependence on redshift. Any global, redshift-independent bias would only change the cross-normalization constant between quasars and Type Ia supernovae, without altering the overall shape of the Hubble diagram. Possible redshift-related effects can be divided in *intrinsic*, i.e. due to a real evolution of the relation between UV and X-ray flux, and *external*, i.e. due to selection effects and/or to incorrect estimate of the UV and/or X-ray flux.

**A. Intrinsic effects in the $L_X$-$L_{UV}$ relation.** The obvious physical effect that would undermine our method, and falsify our results, is a redshift evolution of the relation between the X-ray and UV luminosities. Supplementary Figure 7 shows the redshift evolution of the parameter β which would be needed to explain the difference between our cosmographic fit and the ΛCDM model. Whilst it is formally impossible to rule out such a possibility on a purely observational basis, several considerations suggest it is highly unlikely. We have three main arguments to support the non-evolution of the relation with redshift:

*1. Constancy of the slope of the relation.* We have already shown in previous works[6,7,8] that the slope of the relation remains constant, within the uncertainties, over the whole redshift range probed by our data. This is confirmed with the refined sample used here with an even smaller dispersion than our previous studies (Supplementary Figure 8). Therefore, the redshift dependence should be limited to the normalization of the relation.

*2. Match with supernovae.* The agreement between the Hubble Diagram of Type Ia supernovae and quasars in the common redshift interval (Main Paper Figure 2) shows that no significant evolution can be present up to redshift z=1.4 (unless, for a physical conspiracy, the same effect impact on the supernovae sample in a similar fashion).

*3. Evidence of non-evolving UV and X-ray spectral properties.* Both the X-ray and optical/UV spectral properties of quasars do not show any redshift evolution. In X-rays this can be best seen in the photon index – redshift non-evolution in our sample within the range of photon indexes and redshifts probed (see Extended Figure 9). In the optical/UV the lack of redshift dependence of the main continuum and line features has been shown in the studies of SDSS/HST quasar samples[43,44,45,46,47], and is illustrated by the similarity between the (rest frame) UV spectra of *z*>7 quasars and the composite SDSS quasar spectrum[10]. Once again, our analysis further reinforces the point that no other known quasar property shows a sudden redshift dependence starting from z~1.5, yet we cannot completely rule-out a non-evolving normalization of the $L_X$-$L_{UV}$ relation.

**B. Observational effects.** Possible residual redshift-dependent observational effects may include the presence of dust reddening and gas absorption, the Eddington bias (already discussed above) and cross-calibration issues among the different subsamples. In particular, dust reddening in the UV and absorption in the X-rays may introduce a redshift-dependent systematic effect: moving to higher redshifts, the rest-frame optical/UV and X-ray spectra shift to higher frequencies, where the dust absorption cross-section in the UV is *higher*, and the gas absorption in the X-rays is *lower*. This could lead to underestimate the UV to X-ray ratio, which in turn, based on Equation 1, would imply a higher value of the luminosity distance than what we measured. Supplementary Figure 7 (right axis) shows the average UV extinction needed at each redshift to reconcile the cosmographic fit with the ΛCDM model. To eliminate these biases, we applied strong cuts in optical colours and X-ray slopes as discussed above. We further investigated these issues by analysing how the cosmological results depend on the choice of our filters, and by comparing our flux estimates with those obtained from a complete spectroscopic analysis of selected subsamples. The details of this procedure are outlined in the Supplementary Data. Here we only summarize the main conclusion: while we expect that an improved sample selection and spectral analysis will increase the sample size and decrease the observed dispersion, there is no indication of any significant redshift-dependent observational bias in the final sample.

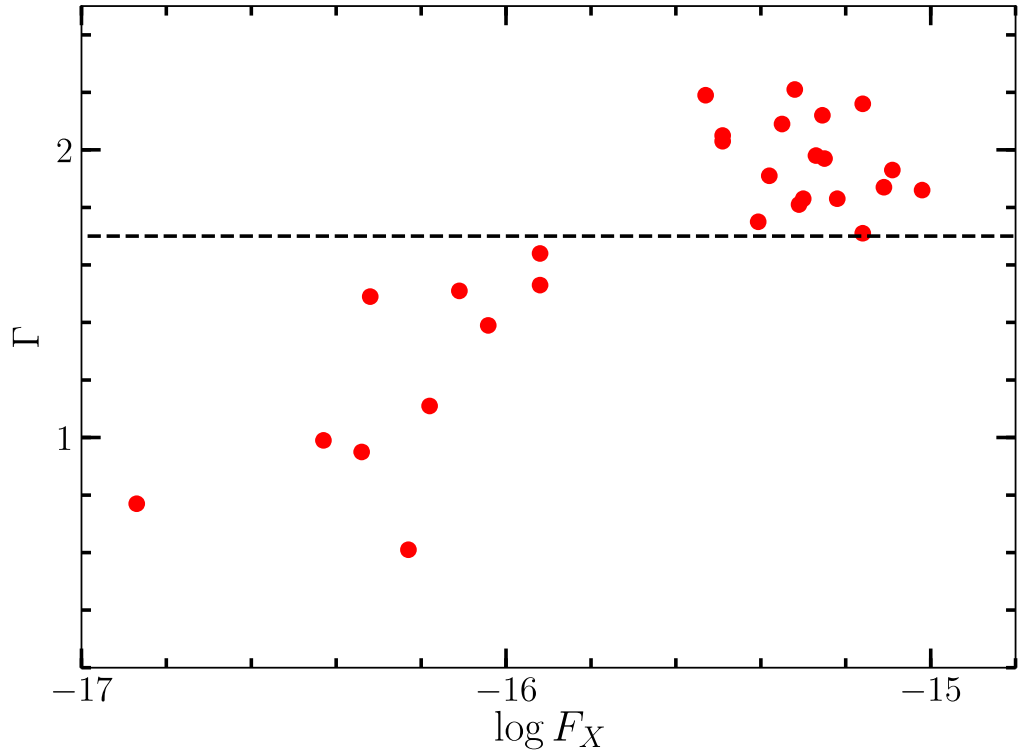

**Supplementary Figure 1:** X-ray photon index as a function of the soft X-ray flux (in erg/s/cm$^2$) for our sample of *z*~3 quasars with dedicated XMM-*Newton* observation. Only quasars above the dashed line are included in the final sample considered in the analysis. Average uncertainties (1σ) on Γ for the bright and faint quasar subsamples are plotted as black points on the right end side of the plot.

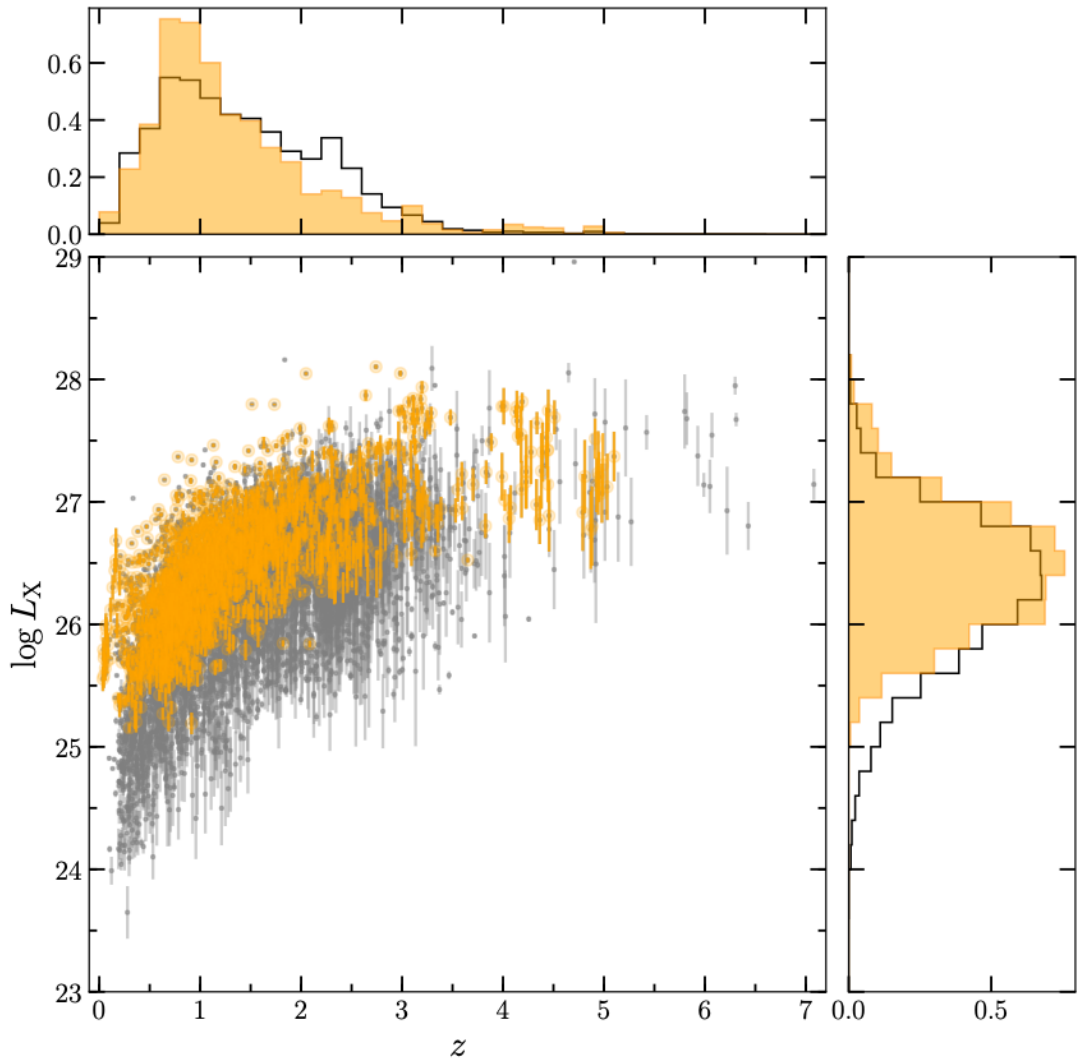

**Supplementary Figure 2:** Monochromatic rest frame luminosity (and 1σ uncertainties) as a function of redshift. The normalized redshift and rest-frame monochromatic 2 keV luminosity distributions for the whole sample of ~7,000 quasars with rest-frame UV and X-ray flux measurements (grey) and for the clean sample of the ~1,600 quasars (orange) selected for the Hubble Diagram. The distribution of both redshift and luminosities are shown in the top and right-end panel, respectively.

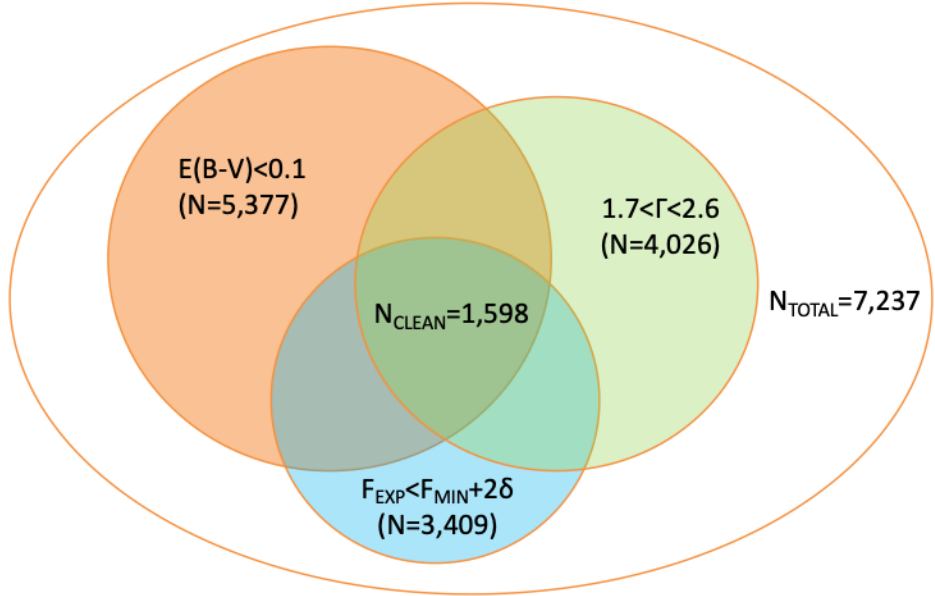

**Supplementary Figure 3:** Summary of the different sub-samples for a given selection criteria.

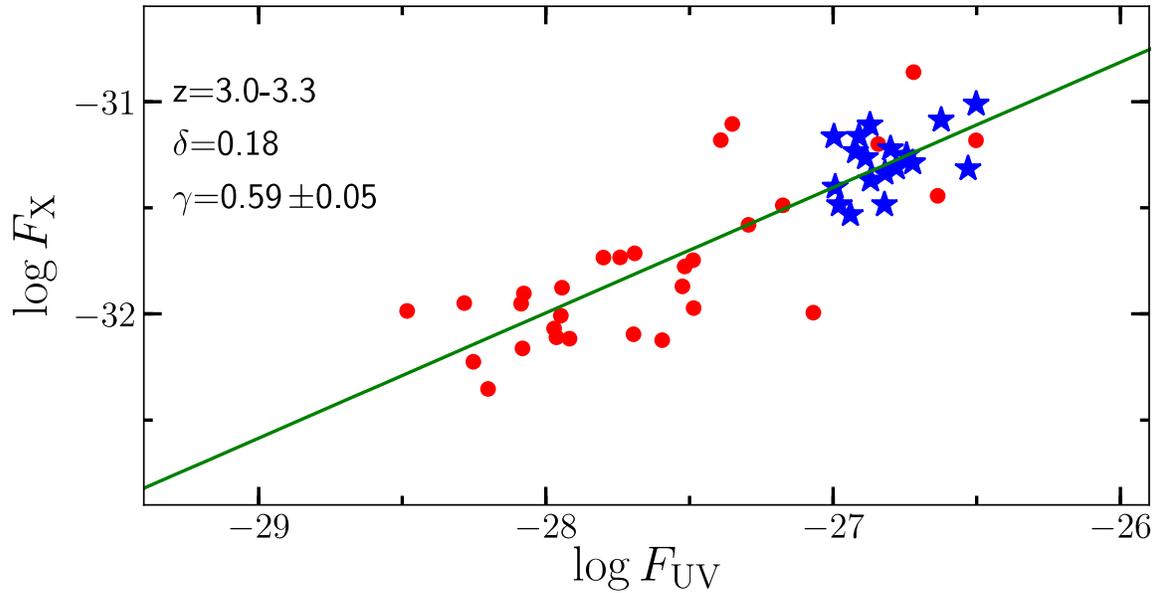

**Supplementary Figure 4:** Rest-frame X-ray (and 1σ uncertainties) to UV flux relation for the sources in our sample within the redshift range 3.0<$z$<3.3. Thanks to the small redshift range, the luminosities can be approximated with the observed fluxes. In this way, the relation is cosmology-independent. The slope is the same as the average one for the total sample, which is dominated by low-z sources. This is a further indication of the absence of any significant evolution of the relation. Blue stars represent the high-quality sample with new XMM-Newton observations.

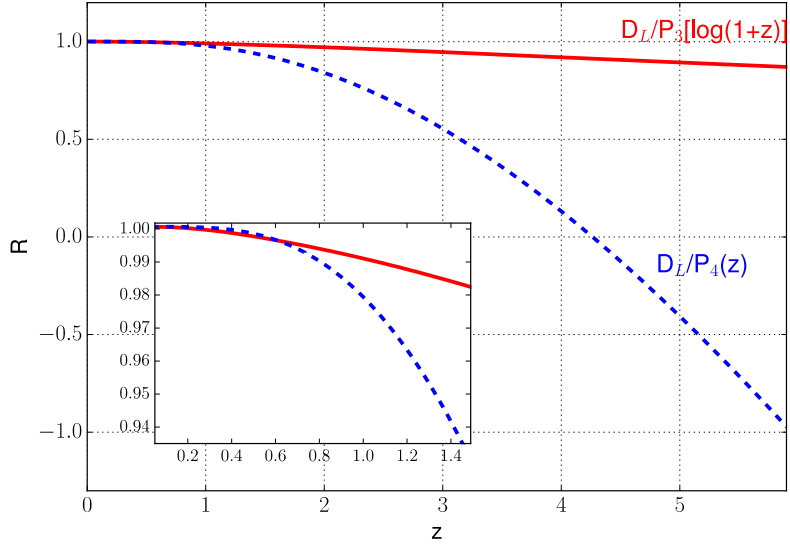

**Supplementary Figure 5:** Ratio between the luminosity distance – redshift relation in a flat ΛCDM model and its approximation with a third order expansion in log(1+$z$) (red continuous line) and a fourth order polynomial expansion (blue dashed line). The small panel is a zoom at low redshifts. The plot shows that the logarithmic expansion is a reasonable approximation of the theoretical values up to high redshift, while the standard polynomial expansion can be used up to z~0.8.

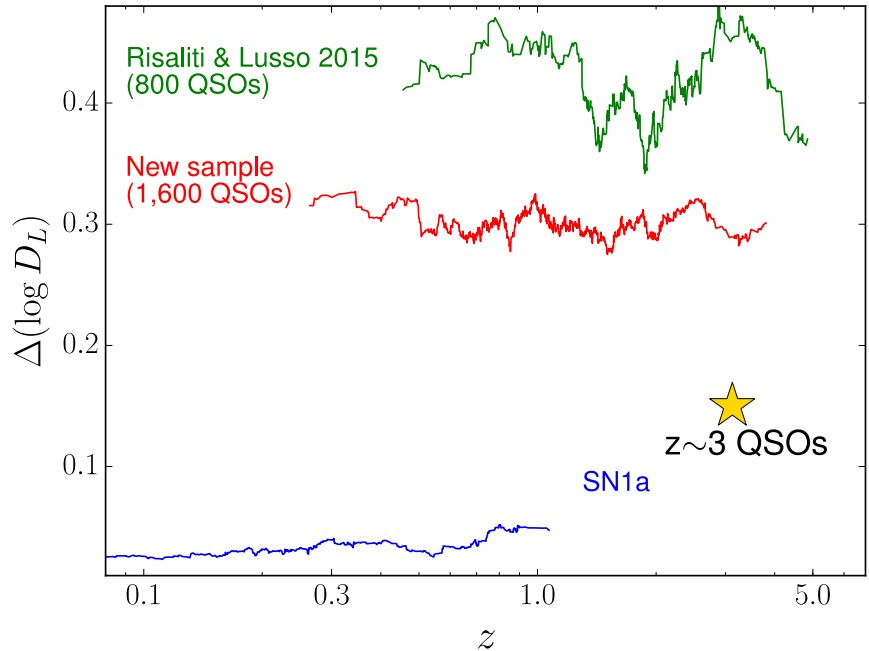

**Supplementary Figure 6:** Scatter on the Hubble Diagram as a function of redshift for the Type Ia supernovae from the JLA sample (blue line), the sample from our first work on the topic where we considered X-ray and UV flux measurements from the literature (green line), and the improved quasar sample presented in this manuscript (red line, Figs. 2 and 9). The star marks the value of the scatter on the Hubble Diagram obtained by considering the quasar sample at $z$=3.0-3.3 with dedicated XMM-*Newton* observations.

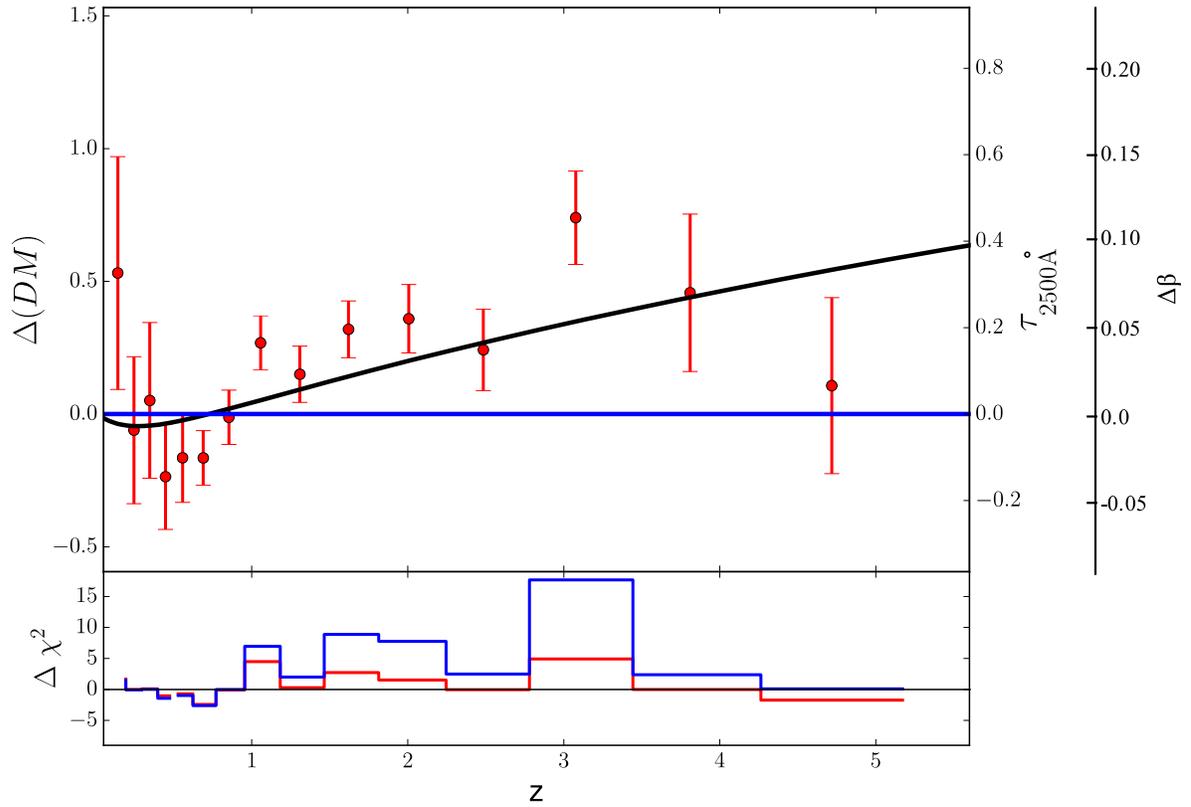

**Supplementary Figure 7:** Difference between the redshift-averaged distance moduli of our quasars (red points and 1σ uncertainties, see the Hubble diagram in Figure 2 on the main paper) and the best-fit ΛCDM model (from a joint fit to SNe and quasars fit, i.e. the dashed line in Figure 2 of the main) as a function of redshift. The black line is the difference between the best fit cosmographic model and the ΛCDM model. The bottom panel shows the $\Delta\chi^2$ with respect to the ΛCDM model (blue histogram) and to the cosmographic model (red histogram). The right y-axes show the same in different units: $\Delta\beta$ is the variation of the parameter β in the $L_{UV}$-$L_X$ relation that would be needed to reconcile the data with the ΛCDM model. Similarly, $\tau_{2500\text{Å}}$ is the optical depth at 2500 Å that would be needed to ascribe the discrepancy to a non-corrected dust extinction on the UV flux.

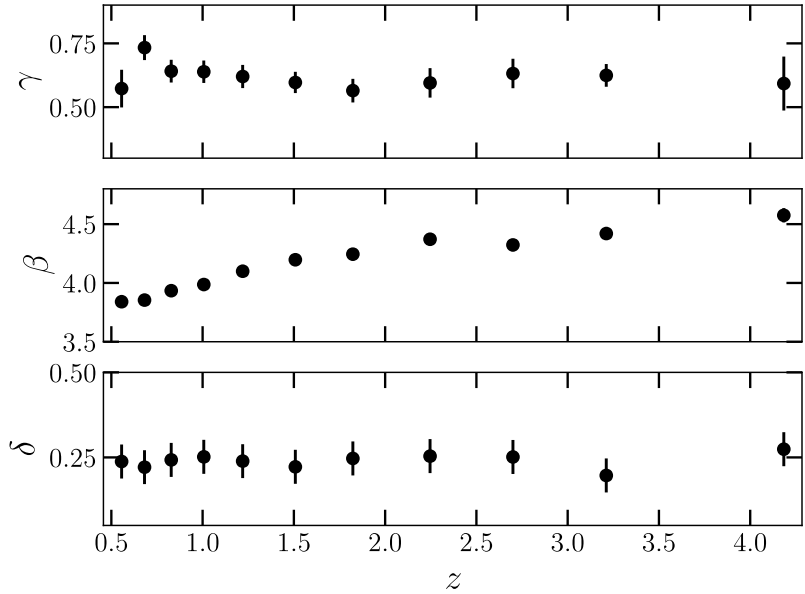

**Supplementary Figure 8:** Evolution of the slope γ, the normalization β, and the dispersion δ of the $F_X$-$F_{UV}$ relation in narrow redshift bins (Δlog z=0.06), where the monochromatic fluxes are normalized to $10^{-27}$ erg/s/cm$^2$/Hz. Error bars represent the 1σ uncertainty on the mean in each bin. The grey solid and dashed lines are the weighted means and 1σ uncertainties of the black points for the slope (γ=0.607±0.05) and the dispersion (δ=0.24±0.05). The red line represents the average slope of ~0.6. As the parameter β is directly proportional to the distance modulus, the trend of β with redshift is simply the quasar Hubble diagram.

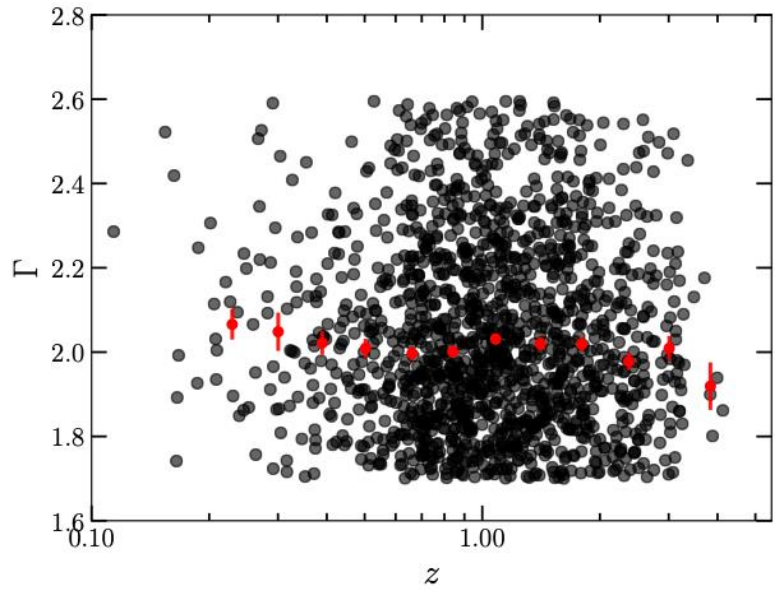

**Supplementary Figure 9:** Distribution of photon indexes as a function of redshift for the clean sample of ~1600 quasars. The red circles represent the mean Γ values (along with 1σ uncertainties on the mean) in narrow redshift bins (Δlog z=0.06).

**Supplementary Material**

In this Supplementary part we discuss the possible systematic effects related to the estimates of the X-ray and UV fluxes, and to the sample selection.
The outline:
1. Measurement of the X-ray fluxes
2. Measurement of the UV fluxes
3. Check of biases in the selection of the sample
4. Check of biases in UV measurements
5. Check of biases in X-ray measurement
6. Checks on the choice of the UV and X-ray frequencies
7. Indications from the new z>3 quasar sample
8. X-ray and UV variability

### 1. Measurement of the X-ray fluxes.

The monochromatic X-ray fluxes at the rest-frame 2 keV ($F_X$) are estimated for the different sub-samples as detailed below. We obtained the $F_X$ values for the 19 quasars in the *Chandra*-Champ and the low-redshift sample (18 AGN) from the respective published papers[20,21].
For the sources in the new z>3 XMM-*Newton* sample and the archival z>4 sample (38 quasars), the $F_X$ values are derived from a complete spectral analysis performed on each individual object. While the results of this analysis will be fully presented in dedicated papers, here we outline the adopted procedure. We fitted the background-subtracted spectra with a power law model, including Galactic photoelectric absorption. We also considered additional components such as Compton reflection and/or additional line of sight absorption, but in all cases the inclusion of these supplementary components was not statistically significant. We thus estimated the $F_X$ values from a simple power law with a Galactic absorption model. The X-ray spectral analysis was performed with the XSPEC 12.9 package, where the rest-frame 2 keV monochromatic flux was explicitly used as a free parameter, together with the photon index of the power law. In this way the errors on $F_X$ are estimated directly, without complications related to the propagation of the errors from measured parameters with non-zero covariance.

For the SDSS/XMM-*Newton* and the XMM-COSMOS samples, which compose the bulk of our quasar sample, we derived the 2 keV fluxes and the relative $\Gamma$ (along with their 1σ uncertainties) from the tabulated fluxes in the 0.5-2 keV band ($F_{SOFT}$) and in the 2-12 keV band ($F_{HARD}$), using a procedure we developed to obtained reliable measurements and errors. The method is outlined as follows.

1. We first simulated a high-quality power law spectrum, assuming average background and calibrations typical for XMM-*Newton*[17], with a photon index $\Gamma=1.7$ (the same used to derive the fluxes in the 3XMM catalogue). We fitted the data in the soft band with a power law parametrized as $f(E)=f(E_0)(E/E_0)^\Gamma$, with $f(E_0)$ and $\Gamma$ as free parameters. We repeated the fit for several values of $E_0$ (from 0.5 to 1.5 keV), and for each case we derived the $f(E_0)$-$\Gamma$ error contour. In general, we obtained inclined contours, due to the non-zero covariance between the parameters $f(E_0)$ and $\Gamma$. However, for the "pivot energy" ($E_0=E_S$), the resulting contours display the main axes parallel to the $f(E_0)$ and $\Gamma$ axes (Supplementary Figure 10). Such $E_S$ value represents the point dividing the soft band in two regions having the same statistical weight, and it is not located at exactly the centre of the energy band because of the dependence of the effective area on the energy. The pivot energy $E_S$ has three relevant properties: 1) the value of $f(E_S)$ is independent from the value of $\Gamma$, implying that our $F_X$ values are insensitive to the specific value of $\Gamma$ assumed in the 3XMM catalogue to derive $F_{SOFT}$. Specifically, in case the "true" photon index of a source deviates from 1.7, the monochromatic flux at $E_S$ derived from $F_{SOFT}$ will be accurate anyway. 2) The covariance between the monochromatic flux at $E_S$ and the photon index $\Gamma$ is zero. 3) The error on the monochromatic flux at the pivot energy, $\Delta[f(E_0=E_S)]$, is the smallest for any choice of $E_0$, and its relative value is equal to the relative error on $F_{SOFT}$, i.e. $\Delta[f(E_S)]/f(E_S) = \Delta[F_{SOFT}]/F_{SOFT}$. For the soft band, we found $E_S=1.05$ keV.

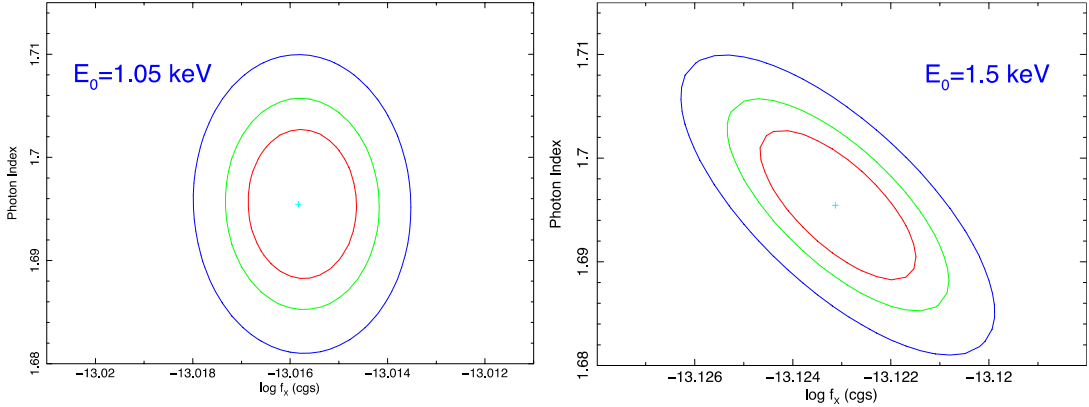

**Supplementary Figure 10:** Contour errors (1, 2 and 3 $\sigma$ for the red, green and blue curves, respectively) for the monochromatic flux $f(E_0)$ and the photon index $\Gamma$ from two fits of a simulated XMM-Newton spectrum in the 0.5-2 keV band. The input model is a power law with $\Gamma=1.7$. In both cases the fitted model is also a power law $f(E)=f(E_0)(E/E_0)^{\Gamma}$, the only difference being the value of $E_0$. The contour on the right panel is the typical one, with a non-zero covariance between the two parameters. The left panel shows the only case with no covariance.

2. We repeated the same procedure for the hard band, 2-12 keV, obtaining a "pivot energy" $E_H=3.1$ keV. We then derived the monochromatic fluxes $f(E_H)$ and relative errors $\Delta[f(E_H)]$ for each source in our sample.
3. For each object, we then considered the power law connecting the two points $[E_S, f(E_S)]$ and $[E_H, f(E_H)]$ to obtain the monochromatic flux at 2 keV (rest frame), and the "true" photon index corresponding to the two measured broad band fluxes $F_{SOFT}$ and $F_{HARD}$. We finally estimated the error on $F_X$ from a simple error propagation, as the errors $\Delta[f(E_S)]$ and $\Delta[f(E_H)]$ are independent.

We tested the results of this procedure by considering a sample of 20 randomly selected objects within the SDSS/XMM-*Newton* and the XMM-COSMOS samples having different observational properties in terms of exposure time, off-axis angle, and background level. We performed a complete spectral analysis of each source, we derived the monochromatic flux at the rest-frame 2 keV and compared these values to the ones resulting from the procedure outlined above. Overall, we found an excellent agreement (within ~1$\sigma$) between the results of the complete spectral analysis and our procedure: the fluxes have no systematic deviation from a 1:1 relation, with a dispersion of the order of the average statistical error on the data points. We conclude that this method provides measurements of both the flux and the error at 2 keV in the most precise possible way, starting from two broad band fluxes, relying only on the assumption that a power law is a good representation of the X-ray spectrum of the source. Moreover, we checked, within the main quasar sample, whether there is any trend between uncertainties on the X-ray fluxes (i.e. luminosities) estimated from X-ray spectral fitting and the number of detected X-ray counts, finding that bright sources naturally tend to have lower flux uncertainties due to the higher count statistics. Yet, as the majority of the quasars within the clean sample have relatively good quality X-ray data, this trend is significantly reduced (as shown in Figure 2 of the main paper).

## 2. Measurement of UV fluxes

To perform our analysis, we utilized the observed continuum flux density values at rest-frame 2500 Å ($F_{UV}$) as compiled by Shen et al. (2011)[18] for SDSS-DR7, which take into account both the emission line contribution and the UV iron complex. These observed flux densities are divided by (1+$z$) to shift these values into the rest-frame. To obtain the rest-frame monochromatic fluxes at 2500 Å for the quasars within the SDSS-DR12 and XMM-COSMOS samples, we used all the available multi-colour information compiled in both catalogues. XMM-COSMOS[22] includes multi-wavelength data from mid-infrared to ultraviolet (UV): MIPS 24 µm GO3 data, IRAC flux densities, near-infrared Y-J-H-K-bands (2MASS and/or UKIDSS), optical photometry (e.g. SDSS, Subaru, CFHT), and near- and far-UV bands (GALEX). Similar broad-band coverage is also available for the SDSS-DR12 quasar catalogue[19]. Observed magnitudes are converted into fluxes and corrected for Galactic reddening by employing a selective attenuation of the stellar continuum[48] with $R_V$=3.1. Galactic extinction is estimated for each object in all catalogues. For each source we considered the flux and corresponding effective frequency in each of the available bands. The data for the spectral energy distribution (SED) computation from mid-infrared to UV (upper limits are not considered) were then blueshifted to the rest-frame and no K-correction was applied. We determine a "first order" SED by using a first order polynomial function, which allows us to build densely sampled SEDs at all frequencies. This choice is motivated by the fact that a single interpolation with a high-order polynomial function could introduce spurious features in the final SED. $F_{UV}$ values are finally extracted from the rest-frame SEDs in the log $\nu$-log $\nu F_\nu$ plane. In the case data do not cover 2500 Å, fluxes are extrapolated at lower (higher) frequencies by considering, at least, the last (first) two photometric data points. We verified our photometric measurements of $F_{UV}$ by comparing them to the spectral ones already published by Shen et al. We find a good agreement between our values and the Shen et al. measurements, with a mean $\Delta \log F_{UV}$= -0.05 and a dispersion of 0.12 dex[6]. We obtained the $F_{UV}$ values for the 19 quasars in the *Chandra*-Champ and the low-redshift sample (18 AGN) from the respective published papers[20,21]. The $z$>3 XMM-*Newton* quasar sample is selected from SDSS-DR7, so we considered the $F_{UV}$ values from Shen et al. Regarding the archival $z$>4 sample (38 quasars), the $F_{UV}$ values are derived from a complete optical-UV spectral analysis performed on each individual object applying a custom spectral fitting code we developed. Our code takes into account the emission line contribution and the UV iron complex in the vicinity of the 2500 Å in a similar fashion as in SDSS.

## 3. Check of biases in the selection of the sample

*Radio bright/jetted quasars.*
It is well established that radio bright (also called radio-loud) AGN have an enhanced X-ray emission mechanism linked to the jets, which can provide an increment in the X-ray emission with respect to radio quiet AGN at similar optical/UV luminosities[49,50]. Radio bright AGN thus occupy a specific region in the $L_X$-$L_{UV}$ relation, namely at high $L_X$ (>$10^{26.5}$ erg/s/Hz). These objects bias the slope of the relation towards steeper values (i.e. $\gamma$>0.7[51]), thus they must be factored out from the main sample.

Radio bright quasars have been excluded from the final quasar sample by using two main methods. The first cleaning has been performed on the SDSS-DR7 sample by excluding all radio emitters with radio loudness (listed in the DR7 quasars catalogue and defined as the ratio between the flux densities at 6 cm and 2500 Å, $R=F_{6cm}/F_{2500Å}$) higher than $10^{52}$. The fraction of radio bright quasars within SDSS-DR7 is on the order of 12%, consistent with previous works in the literature. The SDSS DR12 quasar catalogue does not provide the radio loudness information, we thus computed the $R$ parameter for each quasar in the catalogue following the same procedure adopted by Shen and collaborators for consistency. Specifically, the rest-frame 6 cm flux density is determined from the FIRST integrated flux density at 20 cm assuming a power-law slope of $\alpha_\nu = -0.5$. To further remove powerful radio bright quasars in both the SDSS-DR7 and DR12 samples, we considered the catalogue of SDSS + WISE + 3XMM + FIRST + NVSS quasars published by Mingo et al. (2016, MIXR hereafter)[29]. We also removed all radio loud objects defined as such in the MIXR catalogue (~230 quasars in total). We further cross-matched both SDSS samples to the FIRST/NVSS catalogue released by Mingo et al. (2016)[29], and we excluded all quasars with a 1.4 GHz luminosity higher than $5\times10^{41}$ erg/s (see their Figs. 15 and 16). The majority of radio loud quasars in SDSS-DR7 with $R>10$ also have radio luminosities $>5\times10^{41}$ erg/s[7]. The radio bright quasar fraction within SDSS-DR12 is, again, consistent with previous works in the literature (~13%).

Only 1 radio loud object is present in the new $z>3$ XMM quasar sample, and the XMM-COSMOS sample has been cleaned from radio-loud sources following the technique presented by Lusso et al. (2010)[14]. Radio-loud quasars in the *Chandra*-Champ and the low-redshift samples have been removed following the classification provided in the published papers. The $z>4$ quasars are all radio quiet.
Nonetheless, it is possible that residual jet emission may still contribute in the X-rays for quasars having $R<10$. To quantify a limit on the X-ray emission from radio-quiet sources that is due to jets, we used equation (2) in Worrall et al. (1987)[51]. We find a radio contamination fraction in the X-rays of less than 1% (see also Just et al. 2007[13]).

*Broad absorption line quasars.*
We have excluded all quasars in both the SDSS-DR7 and DR12 catalogues flagged as broad absorption line (BAL). We further checked the sample for additional BALs by cross-matching it with the list of BAL quasars provided by Gibson and collaborators[28]. The BAL classification provided in SDSS-DR7 is mainly based upon the CIV line, whilst the one in the SDSS-DR12 catalogues was carried out with a more systematic approach, also considering other lines such as the low-ionization MgII. Additionally, all the spectra in DR12 have been visually inspected by the SDSS collaboration to ensure the identification of the objects and to identify peculiar features (e.g. broad troughs) that may be missed by the automatized classification. We then compared the BAL classification for the overlapping quasars in the two SDSS catalogues, excluding quasars that have been flagged as such in DR12 but not in the DR7 release. We also visually inspected the optical-UV spectra of the final sample to identify further BALs, but we did not find any within the SDSS coverage. We stress that BALs are known to be X-ray obscured[33,34,35], causing an artificial steepening of the $\alpha_{OX}$-$L_{UV}$ and $L_X$-$L_{UV}$ relations, and they should constitute only ~ 10-15% of the entire quasar population[36]. Given that we excluded quasars with high level of X-ray absorption and checked the slope of the $L_X$-$L_{UV}$ correlation in redshift bins, we are confident that the BAL contamination in the SDSS/BOSS sample is negligible.
BAL quasars in the *Chandra*-Champ and in the low-redshift samples have been removed following the classification provided in the published papers. The optical-UV spectra of the $z>4$ quasars have been visually inspected and none of them show signs of broad troughs in the continuum typically observed in BALs. BALs in the XMM-COSMOS sample are non-trivial to identify as the spectral information varies greatly in both wavelength coverage and signal-to-noise[22]. As such, we considered the following approach. We first selected the XMM-COSMOS AGN that fulfil both optical and X-rays selection criteria and we prioritize over the other quasar samples for overlapping sources, leading to 102 objects remaining out of 542 AGN. Only five out of 102 AGN have no spectral information, so the redshifts are computed from fitting the broad-band SED[53-54]. The measured accuracy on these photo-$z$ is $\sigma_{\Delta z/(1+z)}=0.015$, which is sufficient for our purposes. None of our results are affected by the presence of these 5 AGN with photo-$z$ measurements. We then visually inspected the spectra of those objects for which spectral information was available.

*X-ray undetected objects.*

The fraction of objects undetected in X-rays in the whole sample (i.e. before the filtering process) is of the order of 20% in the SDSS-DR7 subsample, as discussed in a previous 2016 paper[7]. In principle, this may affect the shape of the Hubble diagram of quasars, since more undetected objects are expected at higher redshifts. However, the fraction of undetected sources drops to <2% in the final filtered sample, mainly thanks to the selection on the expected X-ray flux, discussed in the Methods Section in order to remove a possible Eddington bias. In our final clean sample, only sources with an expected X-ray flux significantly higher than the minimum detectable flux for each observation are present. This implies that (a) most X-ray non-detections would be filtered out; (b) the few remaining ones would have upper limits in the X-ray flux higher than the expected flux by at least twice the dispersion (see Methods). Such data have negligible statistical weight in the fitting procedure (and this would be the case even if their fraction were much higher than in our case). We thus conclude that no change in the results would be obtained by including the X-ray non-detections in the statistical analysis.

*Cross-calibration issues*

In order to check for possible cross-calibration issues amongst the different subsamples, we measured the contribution to the dispersion of the distance-redshift diagram for each sub-sample, and we verified that there is no systematic deviation in any of them. This is visually illustrated in Supplementary Figure 11.

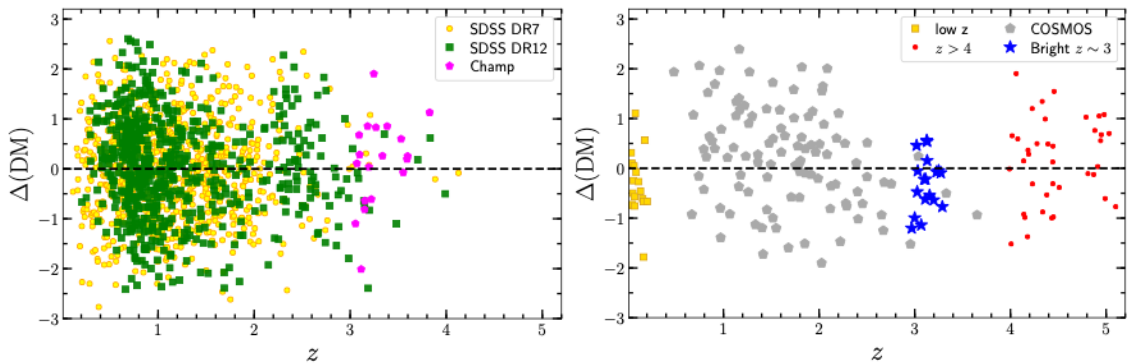

**Supplementary Figure 11:** Residuals with respect to the best-fit, obtained with the cosmographic approach, presented in the Hubble Diagram of quasars in Figure 2. Different colors/symbols are for the different subsamples. The two panels are identical and are duplicated for the sake of clarity.

### 4. Check of biases in UV measurements

In this section we discuss in detail all the possible biases in the UV flux measurements that may affect our cosmological results. We remind that, since the discrepancy between our Hubble diagram and the concordance model is due to a flattening of the Hubble diagram at z>1.5, an overestimate of the UV measurements at high redshift would mean that the true Hubble diagram is even flatter, i.e. that the discrepancy becomes more significant. Instead, a redshift-dependent underestimate of the UV flux would imply higher distances for high-redshift quasars, and hence a better agreement with the concordance model. We split the analysis in two parts, where we discuss (a) possible systematic effects in the fit of the UV spectra, and (b) possible reddening effects.

**(a) Fitting procedure.** The 2500 Å continuum flux values, resulting from the fitting procedure adopted by Shen et al., are computed taking into account the FeII complexes, but not the Balmer continuum, which may produce $F_{UV}$ which are systematically overestimated. Additionally, since we also computed the $F_{UV}$ values from broad-band photometry, the resulting combined sample may also suffer in a systematic offset amongst different sub-samples. We tackle both problems in a two-step approach.

First, we compared the continuum $F_{UV}$ measurements from the tabulated SDSS-DR7 quasar catalogue and the ones computed (along with their uncertainty) from the broad-band SEDs for all the quasars in the SDSS-DR7 catalogue with photometric coverage at the rest-frame 2500 Å, where BALs and radio-loud quasars have been removed (82,215 quasars). This comparison is shown in Supplementary Figure 12. The inset plot of Supplementary Figure 12 presents the average values of the difference between the continuum flux estimated by Shen et al. (2011)[18] and our photometric measurements ($\Delta \log F_{UV} = \log F_{UV,Shen} - F_{UV}$) as a function of the average redshift in each bin. Our $F_{UV}$ estimates are in good agreement with the ones computed by Shen et al., yet there is some residual trend of $F_{UV}$ when we bin the sample in narrow redshift intervals ($\Delta \log z = 0.05$). The solid and dashed lines represent the mean and the error on the mean of the unbinned $\Delta \log F_{UV}$ distribution where the mean is ~-0.06 with a dispersion σ=0.13 in logarithm units. We corrected our photometric $F_{UV}$ values following such pattern as a function of redshift, thus producing *pseudo* spectroscopic flux measurements, for the SDSS-DR12 data. This effect has been also considered in the simulations, where we found that, with large samples (>100,000 objects) that allow more precise measurements of the cosmological parameters, the effect is not negligible. This correction improves the dispersion in the $L_X$-$L_{UV}$ relation by roughly 10-15%. The statistical significance of the discrepancy from the ΛCDM model in the case we consider the photometric $F_{UV}$ values is at the 3.7-3.8σ level, with the same best fit values of the cosmographic parameters, but slightly larger contours due to the higher dispersion.

Second, to minimize systematics on the $F_{UV}$ estimated by Shen et al. due to the missing contribution of the Balmer continuum, we estimated the $F_{UV}$ values by employing our spectral fitting code where we consider, together to the FeII complex, the Balmer continuum[55]. We fitted the SDSS-DR7 spectra for the z~3 quasar sample and we compared our $F_{UV}$ values to the one listed in the catalogue. The $F_{UV}$ values are systematically lower than the ones we computed by a factor of ~0.3 dex. The points in the Hubble diagram are shifted to the bottom by an equivalent amount. Moreover, the dispersion of the distances with respect to their average value (which is a precise proxy of the dispersion in the Hubble diagram, given its flat slope in the $z$=3.0-3.3 range) estimated with our method is δ~0.10, i.e. smaller than that obtained using the Shen et al. values (δ~0.12).

These checks demonstrate that neglecting the Balmer continuum introduces both a systematic bias and an increase of the observed dispersion. Regarding the former effect, we notice that almost all the sources in our sample have UV fluxes estimated directly (as in the SDSS-DR7 sources) or indirectly (as in the SDSS-DR12 ones, through the correction discussed above) with the Shen et al. method. Therefore, if the shift is redshift-independent, it does not change the shape of the Hubble diagram, but only the value of the cross-calibration parameter (or, equivalently, the parameter β in the $L_{UV}$-$L_X$ relation). It is possible to rule out a significant redshift dependence of this bias, and more in general of the errors on $F_{UV}$ measurements, based on the following considerations. The spectroscopic fit is possible in the redshift range z~0.6-2.8, while at lower and higher redshifts only extrapolations are possible. In this interval the spectral analysis is homogeneous, so we do not expect any systematic redshift-dependent difference due to the fitting procedure. Only a physical evolution of the strength of the Balmer continuum may affect our estimates. However, the excellent match between the Hubble diagram of quasars and that of supernovae makes this case extremely unlikely at $z$<1.4. In order to flatten the Hubble diagram at $z$>1.4, we would need a fainter Balmer continuum at $z$>1.4, so that our estimates of $F_{UV}$ are overestimated at lower redshift rather than at higher redshift. This case can be checked through fits of control samples at different z. Our complete spectroscopic analysis of the new $z$~3 quasars shows that the Balmer continuum is still relevant at these redshifts: the complete fit provides $F_{UV}$ values a factor of ~2 lower than those obtained from the Shen et al. method. This is of the same order, or slightly higher, than what we found fitting random objects at lower redshifts. As a consequence, the points at *z*~3 on the Hubble diagram would be shifted to the bottom by Δ(DM)~1.5. In this case they would strongly deviate with respect to all the other points at lower and higher redshift and, more importantly, they would *increase* the discrepancy with the ΛCDM model. Indeed, it is worth noticing that this sample is on average already slightly below the best fit cosmographic model in the Hubble diagram, i.e. even using the Shen et al. fluxes these objects are deviating more than the average from the standard cosmological model. We conclude that a redshift-dependent bias is unlikely to be present and the check on the *z*~3 sample suggests that such bias would in any case shift the results to the "wrong" direction (in terms of consistency with the standard model). An analogous conclusion can be drawn on a possible luminosity-dependent bias, which could also introduce an observational redshift dependence, due to the flux limit of the optical selection of our sample. The sample at *z*~3 has been selected based on the optical brightness, and since the average z is quite high (*z*~3.15) and the redshift interval is small, this sample is also in the high-luminosity tail of the luminosity distribution of our sample (Figure 1 of the main paper). Therefore, a strong decrease of the Balmer continuum at high luminosities is ruled out.

The consistency of the flux estimates at *z*>3 (where the rest frame 2500 Å region moves out of the observed spectrum) has been carefully checked through a control sample, adopting the following procedure: 1) we randomly selected 20 quasars at redshift *z*~2.2 with $F_{UV}$ estimates from Shen et al. At this redshift the rest frame wavelength interval covered by the SDSS spectra is ~1,200-3,000 Å, i.e. including the 2,500 Å continuum plus the MgIIλ2,800Å part and all the continuum down to the Lyα line. We then simulated higher redshift quasars by cutting the high-wavelength end of the control spectra (obviously it is not possible to include the low-wavelength part that would be present at higher redshifts, but the spectrum blueward of the Lyα is irrelevant for the determination of the continuum at longer wavelengths).

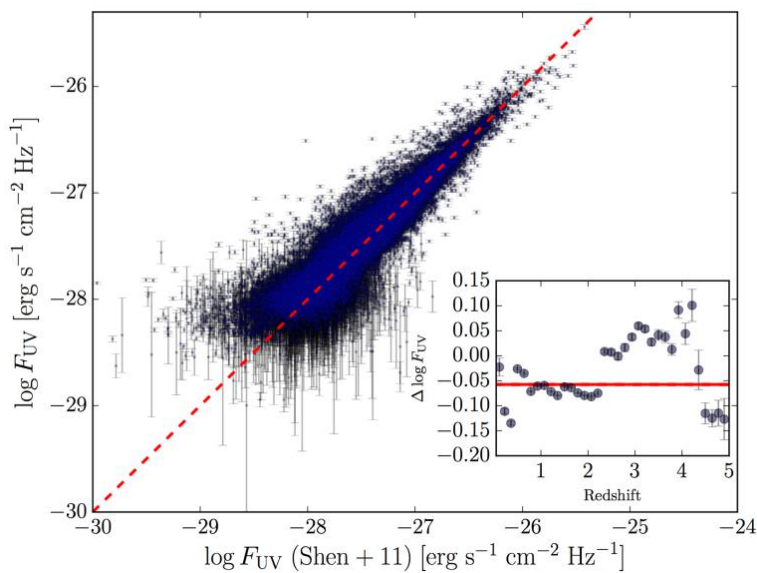

**Supplementary Figure 12:** Comparison of the monochromatic flux at 2500 Å measured from the broad-band SED (along with their 1σ uncertainties) with the flux measured by Shen et al. (2011)[18] through a spectral fit. The dashed red line represents the one-to-one relation. The inset plot shows the average values (along with their 1σ errors) of the difference between the optical flux by Shen et al. and our measurements (Δlog $F_{UV}$ =log $F_{UV,Shen}$ - $F_{UV}$) as a function of the average redshift in each bin (chosen to be narrow enough to exclude further dispersion due to cosmology, Δlog *z*=0.05). The solid and dashed lines represent the mean and the error on the mean of the unbinned Δlog $F_{UV}$ distribution (the mean is ~-0.06 with σ=0.13).

Finally, we estimated the 2,500 Å continuum flux by extrapolating the fits to these artificially redshifted objects, and compared them with the "true" values. We performed a complete fitting procedure up to $z$~4.3. At higher redshifts, we assumed a power law continuum with a fixed slope of 0.79, normalized to the 1,450 Å spectral point, which should be free from major absorption and emission features. At $z$>5.3, where the 1,450 Å point is no longer observed, we tried to fit an average quasar SED to the remaining spectrum in the spectral region redward of the Lyα line. We notice that at λ<1,400 Å we expect a significant absorption by the intergalactic dust[42]. The results of our checks are shown in Supplementary Figure 13, where we compare our estimates with the "true" FUV values for different simulated redshifts.

The conclusions are that (1) it is possible to obtain a precise estimate (i.e. with a dispersion of the order of, or below, 0.05 dex) up to $z$~4.3, and an acceptable, though less precise estimate (dispersion of the order of 0.10-0.13 dex, but no systematic shifts) in the 4.3<$z$<5.3 interval; (2) at $z$>5.3 no acceptable extrapolation is possible with an optical spectrum. Near infrared spectra would be needed. Based on this analysis we included only objects with $z$<5.3 in our final sample.

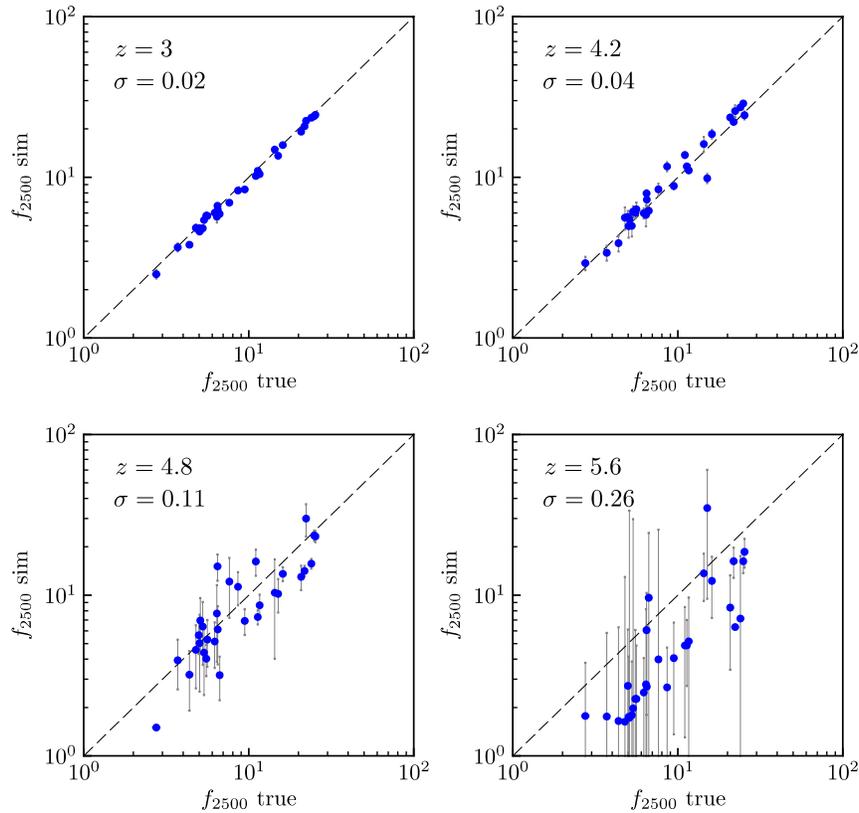

**Supplementary Figure 13:** Comparison between the UV fluxes obtained from the Shen et al. procedure for a sample of z~2 quasars, and the values obtained extrapolating the best fits to the same spectra, performed on the spectral intervals that would be observed if the objects were at higher redshifts. The analysis has been performed for "fake" redshifts from 3.0 to 6.0 in steps of 0.2. Here only four representative cases are plotted, illustrating the main results: the extrapolation is successful for z<5.3, with negligible dispersions at z<4.3, and with dispersions of the order of that of the $L_{UV}$-$L_X$ relation in the 4.3<z<5.3 interval.

**(b) Reddening.** Regarding the dust reddening in the UV, we considered the optical colours derived from the multi-wavelength broad-band SED obtained by collecting all the available photometry from the literature. The procedure of calculating the rest-frame AGN SED from broad-band photometry is described in detail in our previous works[6,7,14]. With respect to our previous work, we improved the selection of blue quasars based on the SED colours by slightly modifying the definition of the $\Gamma_2$ parameter, which is now computed in the interval 1450-3000 Å (as already tested and discussed in Lusso & Risaliti 2016[7]). Moreover, to define a quasar sample where the contamination from dust in the optical is minimal, we computed the predicted $\Gamma_1$, $\Gamma_2$ values from an empirical SED that is representative of the blue SDSS quasar population, i.e. $\Gamma_1 = 0.84$ and $\Gamma_2 = 0.4$, assuming the average quasar SED published by Richards et al. (2006)[56].

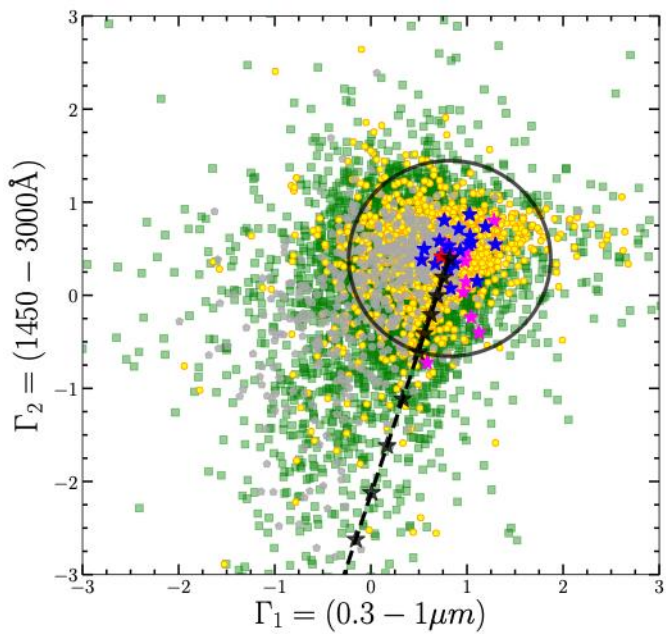

**Supplementary Figure 14:** Distribution of the main quasar sample in a $\Gamma_1 - \Gamma_2$ plot, where $\Gamma_1$ and $\Gamma_2$ are the slopes of a power law in the log $\nu$ – log ($\nu F_\nu$) plane in the 0.3–1 μm and 1450-3000 Å intervals, respectively. The typical error for $\Gamma_1$ and $\Gamma_2$ is of the order of 0.1. The red star represents the intrinsic quasar SED (i.e. $\Gamma_1 = 0.84$ and $\Gamma_2 = 0.4$) published by Richards et al. (2006)[56]. The dashed black line is obtained by assuming increasing dust extinction following the extinction law of Prevot et al. (1984)[57] with E(B-V) values in the range 0.02-0.1, with a step of 0.02 and 1450-3000 Å with a step of 0.5 (marked with black stars). Green squares: SDSS DR12; yellow circles: SDSS DR7; blue stars: z~3 quasars from our XMM Large Programme (blue: $\Gamma$>1.7 "Norm" quasars, magenta: $\Gamma$<1.7 X-ray "Weak" quasars, red: radio loud); grey pentagons: XMM-COSMOS AGN.

However, the values for the optical-UV slopes are likely to vary from the empirical ones in the case of accretion disc (AD) models, which depend upon black hole masses and accretion rates. To demonstrate that our procedure of identifying red quasars is not in contradiction with thin AD models, we computed the optical slopes, $\Gamma_1$ and $\Gamma_2$, for a set of AD models with $M_{BH}=10^8$ and $10^9$ $M_{sun}$, and accretion rates (defined as $L_{bol}/L_{EDD}$, the so-called Eddington ratio) of 0.1, 0.3, and 1. For each combination of black hole mass and Eddington ratio, we also considered the two extreme cases of a Kerr (spin = 0.998) and a Schwarzschild (spin = 0) black hole in order to probe the widest range of possibilities. The $\Gamma_1$ and $\Gamma_2$ values range from ~1.0 to 1.4 and ~0.4 to 0.8, respectively.
We then tried different filters based on the difference with the expected colours of a pure thin AD disc, which can be then translated in a range of reddening, E(B-V), values. We finally produced a distance-redshift diagram for each choice of this filter.

We found that, for the bulk of our sample selected from the SDSS catalogs, the results are insensitive to the colour cut. We interpret this result as the indication that the colour distribution of SDSS quasars is mostly due to intrinsic differences, and not to dust reddening.

The only observed effect is an increase of the total dispersion in the $L_X$-$L_{UV}$ correlation, thus in the distance-redshift relation, when we relax the colour cut to increase the statistics (i.e. we include quasars with a threshold in reddening of E(B-V) < 0.2 and 0.3).

A possible, small fraction of dust-reddened quasars can still be present in the sample, but since they are also absorbed by gas in the X-rays, they are efficiently removed by the other filters on the X-ray spectrum. The increase of the dispersion for strongly deviating objects, without any systematic change of the distance-redshift diagram, is most likely due to incorrect UV flux measurements. Based on these findings, we included in the final sample all the quasars with optical colours corresponding to E(B-V) < 0.1[6,7]. In Supplementary Figure 14 we present the $\Gamma_1$ - $\Gamma_2$ for the main sample (colour coded as described in the caption). The typical error for $\Gamma_1$ and $\Gamma_2$ is of the order of 0.1. The central black star represents the intrinsic quasar SED as estimated by Richards et al. (2006)[56]. The dashed line is obtained by assuming increasing dust extinction following the extinction law of Prevot et al. (1984)[57]. Stars along the dashed line represent increasing reddening, with E(B-V) values from 0.02 to 0.3.

## 5. Check of biases in X-ray measurements

The possibility that our cosmological results are affected by redshift-dependent biases in the estimates of X-ray fluxes can be easily ruled out with high confidence. Given the relation between distances and X-ray fluxes based on the $L_{UV}$-$L_X$ relation (i.e. $\log(D_L) \propto \log(F_X)$, see Equation 1 in the Methods), the discrepancy between the Hubble diagram of quasars and the concordance model would *increase* if the X-ray fluxes were *underestimated*. The method to compute the X-ray fluxes, described in the Methods, relies on the assumption that the X-ray spectrum is well reproduced by a power law. According to a large amount of observational evidence, this is an excellent approximation for the continuum shape, when absorption is absent: possible reflection from circumnuclear material and the soft excess are the main additional components to be considered.

The modification of the spectrum due to a reflection component is always small, especially in quasars, where it is rarely observed, and most of its effect is included in a slightly softer effective photon index. Simulations computed with the XSPEC software show that the error on the flux at 2 keV due to neglecting this component is at most of a few per cent, i.e. totally negligible with respect to the dispersion of the $L_{UV}$-$L_X$ relation (see Supplementary Figure 10). The contribution of the "soft excess" below 1 keV can be relevant, as demonstrated by the spectral analysis of local type 1 AGN[39], and it may affect our estimate of the 2 keV monochromatic flux for objects at $z<1$, since the procedure outlined in the Methods uses the whole 0.5-2 keV band. Our simulations show that - fitting with a single power law - a soft excess+power law spectrum leads to an overestimate of the 2 keV flux by ~10-30% at $z=0$, depending on the strength of the soft excess (the simulations covered the range observed in local type 1 AGN). The consequence of this bias is an underestimate of the distances at low redshift. Though this effect can be better treated through a more careful analysis for the low-redshift objects (either performing a complete spectral analysis or neglecting the 0.5-1 keV part of the spectrum), we note that it cannot significantly affect our results, being limited in the redshift range where the cosmological information is dominated by supernovae. Moreover, the excellent match between the Hubble diagram of supernovae and that of quasars in the common redshift range suggests that this effect must be small compared with the dispersion. Uncertainties on the X-ray fluxes/luminosities scale at the zeroth order with the number of detected X-ray counts and this is especially true for the serendipitous quasar sub-sample, i.e. SDSS-3XMM. To check the accuracy of our measured fluxes, we computed the average trend of the net counts as a function of both redshift and rest-frame monochromatic UV luminosities. Supplementary Figures 15 presents the distribution of the net counts as a function of redshift (left) and $L_{UV}$ (right) and the medians in narrow redshift and $L_{UV}$ bins for the main and the clean quasar samples. Both the main and the clean quasar sample display very flat slopes as a function of $L_{UV}$, indicating that there is no systematic bias. There is a decrease of the net counts as a function of redshift in the case of the main sample (i.e. a factor of about 3 from redshift 0.8 to 3), while the clean sample is much flatter (in the same redshift range). This check again rules out any redshift-dependent biases in the estimates of X-ray fluxes.

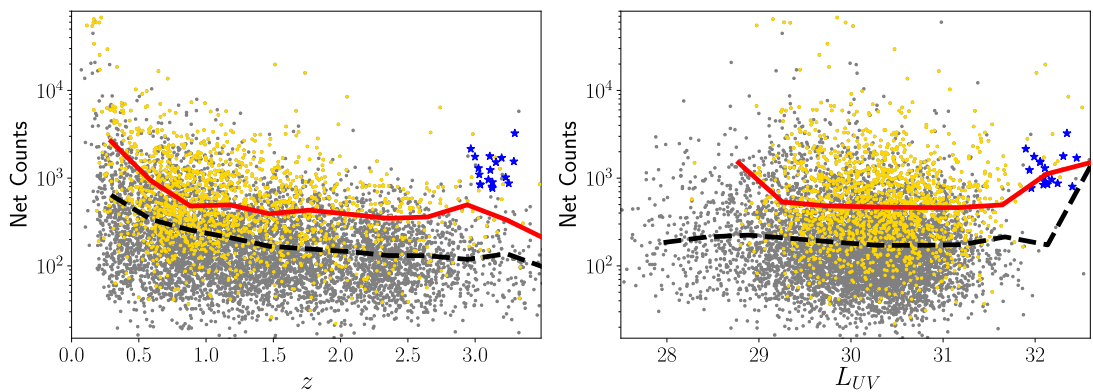

**Supplementary Figure 15:** Distribution of the net counts as a function of redshift (left panel) and the rest-frame monochromatic UV luminosity at 2500 Å (right panel). The grey and gold points represent the main and the clean quasar samples, respectively. Here we only considered serendipitous X-ray observations (i.e. SDSS-3XMM data). The red solid and black dashed lines are obtained by calculating the median of counts in narrow redshift and UV luminosity bins. The z~3 XMM sample (blue stars) is shown as a reference but not used in the averages.

*Choice of the photon index cut.* The most serious issue that could affect the precision of our flux estimates is gas absorption. Previous studies of large AGN surveys[38] show that about 25% of optically selected type 1 AGN present X-ray absorption in excess of the Galactic value (which is taken into account in our analysis). If not corrected for, this absorption leads to an underestimate of the X-ray flux, and an overestimate of the distance. Since absorption mostly affects the low energy part of the spectrum, this bias is expected to be more relevant at low redshift. Therefore, the global effect on the Hubble diagram will be a decrease of the ratio between high-redshift and low-redshift distances, i.e., qualitatively, the effect leading to the discrepancy with the concordance model. In order to remove this bias, we applied a selection based on the photon index. We have tried different cuts in photon index, starting from $\Gamma>2$ down to $\Gamma>1$, in steps of $\Delta\Gamma=0.1$, and we performed a full analysis of our sample in each case. This is done by also including the analysis of the $F_X$-$F_{UV}$ correlation in small redshift bins, and of the Hubble diagram. We did not observe any change in the shape of the Hubble diagram and in its dispersion by considering values from $\Gamma>2$ to $\Gamma>1.5$. In each case we also computed consistent measurements of the cosmological parameters (with decreasing uncertainties, thanks to the increasing number of sources in the sample). When we included sources with $\Gamma<1.5$, we observed a flattening of the Hubble diagram, as expected if absorbed sources start to be included in the sample. Based on the X-ray spectra of local type 1 AGN, we decided to consider objects with $\Gamma>1.7$. With respect to the $\Gamma>1.5$ one, we have ~300 sources removed from the sample, and the statistical significance of the deviation from the concordance model drops from 4.5σ to 4σ. However, we believe this conservative cut is more appropriate for a preliminary work on this subject, where the reliability of the results is more important than the sample statistics.

## 6. Checks on the choice of the UV and X-ray frequencies.

All the analysis has been performed using the monochromatic fluxes at the rest-frame 2500 Å and 2 keV. This choice has been adopted since the first works on the UV to X-ray luminosity relation in quasars[12], and it was mainly based on observational considerations: the UV flux corresponds to the centre of the V band at $z=1$, and the X-ray flux is at the centre of the 0.5-2 keV band also at $z=1$.

It is obviously relevant for our work to revise this choice, keeping in mind the following physical and observational considerations:
1) Regardless of the specific physical mechanism driving the observed relation, it is natural to expect that the physical link must be in the synergy between the wide-band emissions from the UV-emitting accretion disc and the X-ray emitting corona. In this respect, broad-band fluxes may be a better choice for our analysis. Regarding monochromatic fluxes, we expect that the best choices are at the wavelengths which are more closely related to the global emission of both the disc and the X-ray corona.
2) It is important to quantify the impact of a non-optimal choice of the wavelength/energy both in terms of dispersion and possible biases. For example, a monochromatic flux which is not fully representative of the broad-band emission may increase the dispersion, and therefore decrease the statistical significance of our results. Instead, a monochromatic flux with a strong redshift dependence on extinction/absorption, may introduce a bias in our Hubble diagram, invalidating our results. It is clear that the latter effect is more dangerous that the former, and the priority must be to avoid any redshift-dependent bias.

Based on the above general considerations, we revised the choice of the monochromatic fluxes at 2500 Å and at 2 keV, as described below:

**1. UV flux**. The average SED of a blue quasar shows an increase in terms of total fluxes (i.e. $\nu F_\nu$) down to wavelengths $\lambda\sim 1,000$ Å. It is difficult to determine the exact wavelength of the peak emission, due to the high extinction beyond the Lyman limit, however it is probably not much beyond this limit depending on the luminosity/black hole mass[44]. On the redder side of the optical/UV SED, the contribution of the host galaxy becomes non-negligible for luminous quasars only at $\lambda>4000$-$5000$ Å. Therefore, the optimal choice for our study based on physical considerations, is a UV flux between ~1,000 Å and 4,000 Å, with the blue side dominating the broad-band emission. Observationally, we must exclude the 1,000-1,400 Å region, where absorption due to the intergalactic medium starts to be important[44]. In the remaining 1,400–4,000 Å (rest-frame) range, the relevant aspects to be considered are *(a)* the peak of the accretion disc emission is in the low(blue)-wavelength region; *(b)* dust extinction is higher at bluer wavelengths; *(c)* the redshift range where most of the cosmological information resides is between $z\sim 1.5$ and $z\sim 3$. At lower redshifts the statistics is dominated by supernovae, while at higher redshift the poor data quality only allows weak constraints on the cosmological parameters. The 4,000-10,200 Å observed interval of the SDSS spectra corresponds to the rest frame 1,600-4,100 Å at $z=1.5$ and 1,000-2,550 Å at $z=3$. The common spectral range is 1,600-2550 Å. In order to obtain reliable flux estimates, we must choose a wavelength range within this common interval.

The choice of the 2,500 Å wavelength is therefore justified as the reddest point (and therefore the least affected by dust extinction) in the optimal spectral range. An alternative choice is the bluest point, which is more affected by dust extinction, but is closer to the emission peak of the accretion disc. In order to explore this possibility, we performed a complete analysis of the full SDSS sub-sample adopting the monochromatic flux at 1,450Å as UV flux. This is the shortest wavelength without significant intergalactic medium absorption, and it is almost unaffected by narrow emission features, making the continuum determination independent from the details of the spectral fit (this is not the case for the 2,500 Å flux, where the Fe II component makes the determination of the continuum level more complex, as discussed in the previous Sections). The results are shown in Supplementary Figures 16 and 17. The $F_{UV}$-$F_X$ relation in small redshift bins is consistent with the one found adopting the 2,500 Å flux, and the distance estimates are also in full agreement. The dispersion in the Hubble diagram is the same as in the main sample. These results demonstrate that there is no advantage in choosing a bluer wavelength for the estimate of the UV flux. In future works, we will explore the $L_{UV}$-$L_X$ relation and its cosmological application adopting a broad band UV flux. This requires a complete spectroscopic analysis of our sample, which is beyond the scope of this paper. Based on the checks on the 1,450 Å flux, however, we expect a similar result, with, at best, a slightly smaller dispersion.

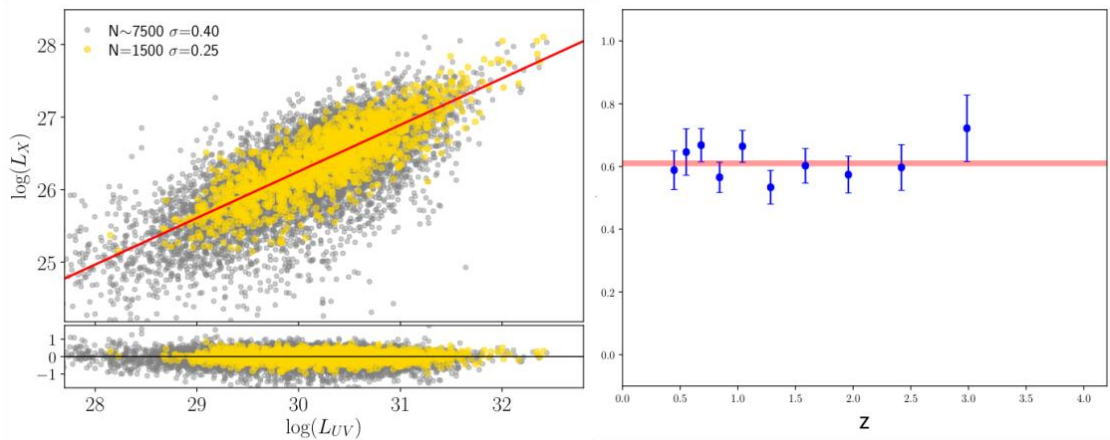

**Supplementary Figure 16:** Results of the analysis of the X-ray to UV relation with the UV flux estimated at 1,450 Å. Left panel: fit of the $L_X$-$L_{UV}$ relation assuming a standard cosmological model. Right panel: best fit slopes of the relation fitted in small redshift bins. In both cases the results of the fit are fully consistent with those found using the 2,500 Å flux.

**2. X-ray flux.** Analogous considerations as made for the UV flux are valid for the choice of the optimal X-ray flux. Our estimate of the 2 keV monochromatic flux for the majority of the sources in our sample is based on the flux values listed in the 3XMM catalogue, as described in detail in the Methods. We briefly summarise our two-step procedure to measure the X-ray fluxes. First, we derive the parameters of a power law spectrum from the 0.5-2 keV and 2-12 keV observed EPIC fluxes, then we use this power law to derive the 2 keV monochromatic flux. Here we discuss the appropriateness of both steps.

- The power law is determined from two spectral points chosen as the "weighted pivot points" in the two bands. The main additional features observed in AGN spectra are a soft excess at about 1.0-1.5 keV, and a reflection component including a flat continuum and a prominent iron Kα emission line. While the soft excess can be dominant over the power law below 1 keV, the contribution of the reflection component is usually negligible up to ~20 keV. Since the effective area of XMM-*Newton* peaks in the 1-2 keV range, and rapidly decreases at higher energies, the hard band is dominated by the counts at the low end of the energy interval. This is clear from the value of the pivot point in the hard band, which is at an energy as low as 3.1 keV. As a consequence, the effective spectral range used to determine the parameters of the power law is ~0.5-5 keV (observed frame) Considering the *z*=1-3 spectral range, the observed interval corresponds to 1-10 keV at *z*=1 and 2-20 keV at *z*=3 in the rest frame. In both cases, the rest-frame interval is expected to be dominated by the power law component. We conclude that the approximation of the X-ray spectrum as a power law is well motivated. We point out again, as discussed in the Methods, that we checked this procedure with several random sources for which we performed a complete spectral analysis, and we did not find significant differences in the estimates of the fluxes and of the spectral parameters.
- Once the power-law fit described above has been validated, the question remains on whether the rest-frame monochromatic flux at 2 keV is appropriate to represent the X-ray emission. In general, we expect that, regardless of the details of the physical mechanism linking the accretion disc and the corona, the physical relation is between the broad-band disc and the X-ray coronal emissions, so a broad band X-ray flux may be a better choice to investigate the $L_X$-$L_{UV}$ relation. This requires a re-analysis of the whole sample, which we plan to do in a future work. However, a qualitative check demonstrates that the improvement in terms of the dispersion of the relation must be small, i.e. we can repeat the argument used to determine the "weighted pivot points" in the 3XMM observed bands (see Methods) on the power law derived from the 3XMM fluxes. A "pivot energy" must exist where the monochromatic flux is independent from the photon index.

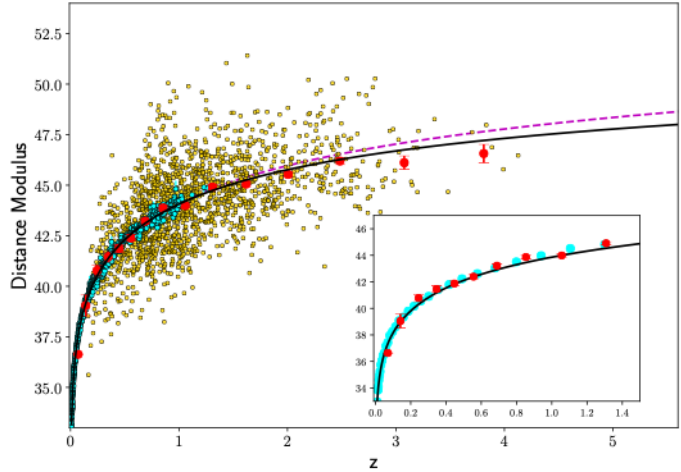

**Supplementary Figure 17:** Hubble diagram, analogous to Figure 2 of the main paper, for the sample of SDSS quasars where the UV flux has been estimated at 1,450 Å. The results are fully consistent with those obtained with the main sample based on the 2,500 Å flux.

The monochromatic flux at this energy has the smallest relative error in the whole energy band. This error is equal to the relative error on the total broad-band flux. Choosing this monochromatic flux or the broad-band one should therefore be equivalent. The determination of this pivot point is not trivial, because it depends on the physical extension of the power law component, (while, in the previous application, it was related to observational parameters, i.e. the extension of the 3XMM spectral bands). These limits are not well known in the low-energy end (due to the presence of the soft excess and to the lack of observations below 0.2-0.3 keV), nor in the high-energy end (the energy of the exponential cut-off is still poorly determined). We can however adopt an observational approach as follows: if $F_C$ and $E_C$ are the flux and the energy at the pivot point, they are related to the energy and monochromatic flux at 2 keV as: $F_{2keV}=F_C(E_{2\,keV}/E_C)^{(1-\Gamma)}$, where $\Gamma$ is the photon index obtained in the first step of the procedure. If the tightest relation is between $L_{UV}$ and $L_{2\,keV}(E_C)$, then from: $\log(L_C)=\gamma\log(L_{UV})+\beta$ we obtain

$$\log(L_{2\,keV})=\gamma\log(L_{UV})+\beta+(1-\Gamma)(\log(E_{2\,keV}/E_C).$$

In other words, we expect to observe a linear dependence on $\Gamma$ of the $L_{2\,keV}$ - $L_{UV}$ relation.
We repeated our fits of the relation adding an extra linear term containing $\Gamma$, in order to check this dependence. We both performed the fits of the $L_X$-$L_{UV}$ relation for the whole sample, assuming a standard cosmology, and of the $F_X$-$F_{UV}$ in small redshift bins, without any assumption on the cosmological model. In both cases, we did not find a significant dependence on Γ (the coefficient is consistent with zero within a 2σ confidence level). We conclude that the 2 keV monochromatic luminosity is already maximally representative of the X-ray emission, and that changing the energy at which the X-ray fluxes are calculated would not decrease the observed dispersion of the relation. This result may be due to the fact that the 2 keV energy is close to the pivot point of the X-ray spectrum. For example, if the physical extension of the power law component is between $E_{MIN}$=0.1 keV and $E_{MAX}$=40 keV, and Γ=2, then $E_C \sim (E_{MIN}E_{MAX})^{0.5} \sim$ 2 keV.

Since a similar dependence holds between $F_{2\,keV}$ and the broad-band flux (the relation is exactly linear in log units only if Γ=2, but the deviations in the range Γ=1.7-2.3, where most quasars lie, are negligible), we would not find any improvement using the broad band flux either.

## 7. Indications from the new *z*>3 quasar sample.

Our new sample of *z*=3.0-3.5 quasars with dedicated XMM-*Newton* observations is a powerful test for many of the effects (both physical and observational) discussed above. We already discussed most of the checks based on this sample throughout the Methods and Supplementary Sections. Here we summarize and expand these arguments.

The sample has been selected based on the blue optical colours, with no evidence of dust extinction, and on its high optical flux. Given its narrow redshift range, this is almost equivalent to a luminosity selection. Indeed, based on the standard virial indicators used by Shen et al. (2011)[18], these quasars have black hole masses in excess of $10^9$ $M_{SUN}$ and Eddington ratios close to 1. A complete analysis of the UV spectra will be presented in a forthcoming paper. The X-ray properties of the sample are illustrated in Supplementary Data Figure 8, where we show the X-ray flux-Γ distribution. The same filter on the photon index we applied to the main sample (Γ>1.7) produces a clear separation between X-ray weak, flat objects and X-ray bright, steep ones. The first group is filtered out from our sample, but it is nevertheless quite interesting for its X-ray and optical properties and will be discussed in a forthcoming dedicated paper.

All the 18 objects in the X-ray bright group are instead selected in the main sample.

Given the selection criteria, and the dedicated X-ray observations, it is unlikely that any systematic effect due to the selection process affects this sample. It is therefore interesting to compare the results on this "optimal" sub-sample with those on the whole sample:

1) The average position in the Hubble diagram of the new objects is slightly below the global best fit (Figure 2 of the main paper). Since these sources have been carefully checked for dust extinction, this confirms that the flattening of the Hubble diagram at high redshifts with respect to the prediction of the standard cosmological model, is not due to a redshift-dependent extinction.

2) A similar consideration can be applied to the correction for the Eddington bias: these sources are obviously not affected by such a bias which, according to our tests and simulations, causes a flattening of the Hubble diagram. It is therefore unlikely that the main sample is significantly affected by this bias as it is slightly above the new points in the Hubble diagram.

3) The new sample has the smallest dispersion observed so far for any sub-sample (only 0.12 dex, see Supplementary Data Figure 3). This clearly demonstrates that most of the observed dispersion is not intrinsic, but due to residual uncertainties in the measurement process. This is a strong validation of our method and, at the same time, the indication that a significant improvement is possible with a more detailed (possibly, source-by-source) spectroscopic analysis.

4) The X-ray to UV luminosity relation has been checked in the *z*=3.0-3.3 interval with high precision, and without any assumption on the cosmological model, given the small redshift interval. No evolution of the slope has been found.

## 8. X-ray and UV variability

We believe that work still needs to be done regarding the effect of the X-ray and UV variability on the relation, which we will present in a forthcoming publication. Variations in the UV brightness are on the order of about 10% (i.e. 0.04 dex in logarithmic units) on time scales of months to years[58-59]. The X-ray variability[60] is on the order of 5% on long time scales at high luminosity and somewhat larger at lower luminosity. It is well known that the UV and X-ray variability are not correlated on short timescales (e.g. NGC5548[61]), so the *intrinsic* variance on the relation could be even lower than 0.1 dex. We cannot compute a one-to-one correction for all quasars within the sample due to the lack of data. Regarding the X-ray variability, we have shown that its effect is of increasing the dispersion of the relation, but it does not change the slope of the relation[7]. This has also been confirmed by previous analyses using simultaneous X-ray/UV data: there is no systematic or significant change in the slope of the $L_X$-$L_{UV}$ relation when using simultaneous data-sets[21-62-63]. The slopes of the $\alpha_{OX}$-$L_{UV}$ relation measured in all previous works in the literature are consistent within the 2σ statistical level. Although in the case of low fluxes, X-ray and UV variability may bias our data towards brighter states, as discussed at length in both Methods and Supplementary sections, both X-ray and UV variability have the only effect of producing higher uncertainties on the final computation of the parameters, without introducing any major systematics. This increase of dispersion is undesirable, given that, ultimately, quasars have the potential of being precise distance estimators. Nonetheless, in this manuscript, we prefer to give priority to measurement robustness.